\documentclass[a4paper,11pt]{article}

\usepackage[latin1]{inputenc} 
\usepackage[T1]{fontenc} 
\usepackage{lmodern} 
\usepackage[a4paper, left=2cm, right=2cm, top=2cm, bottom=3cm]{geometry} 
\usepackage[english]{babel} 

\usepackage{amsfonts,latexsym,amssymb} 
\usepackage{mathrsfs,amsmath}
\usepackage{euscript} 
\usepackage[all]{xy}
\usepackage{placeins}
\usepackage{ctable, multirow}
\usepackage{graphicx}
\DeclareGraphicsExtensions{.eps}

\usepackage{natbib} 

\newlength{\defbaselineskip}
\newcommand{\Proof}{{\noindent {\bf Proof:   }}}
\newcommand{\preuve}{{\noindent {\bf Proof   }}}
\newcommand{\EndProof}{{\hfill $\Box \qquad $ \endtrivlist}\par }

\renewcommand{\1}{ \mbox{ \usefont{U}{bbm}{m}{n}  \selectfont 1}}

\newcommand{\Indic}[1]{ \1_{#1}}

\renewcommand{\le}{\leqslant}
\renewcommand{\ge}{\geqslant}

\newcommand{\marque}{}
\newcommand{\rien}{}
\newcommand{\signesmarge}[1]{\newlength{\tailleetoile}\settowidth{\tailleetoile}{#1} \renewcommand{\marque}{ #1 }
\renewcommand{\rien}{ \makebox[\tailleetoile][]{} }}
\newcommand{\set}[1]{\left\{#1\right\}}
\newcommand{\N}{\ensuremath{\mathbb{N}}}

\renewcommand{\phi}{\varphi}

\renewcommand{\P}{{ \ensuremath{\mathrm{P}} }}

\newcommand{\Proba}[1]{{ \P\left[ {#1} \right] }}

\newcommand{\E}{{ \ensuremath{\mathrm{E}} }}
\newcommand{\Esp}[1]{\E \left[{#1}\right]}
\newcommand{\Var}[1]{\ensuremath{\mathrm{V}} \left[{#1}\right]}
\newcommand{\sachant}{\, \vline \,}

\newcommand{\ordre}[1]{^{({#1})}}

\newcommand{\og}{"}
\newcommand{\fg}{"}

\usepackage{color}
\newenvironment{modification}{}{}
\renewenvironment{modification}{\color{black}}{}
\newcommand{\modif}[1]{{\begin{modification}#1\end{modification}}}

\newcommand{\bino}[2]{{\modif{\mathrm{C}_{#1}^{#2}}} }
\newcommand{\C}{{\modif{\mathscr{C}}} }
\newcommand{\Yt}{Y_t}
\newcommand{\coeff}[3]{#1_{#2, \, {#3} }}
\newcommand{\lambdat}[1]{\coeff{\lambda}{#1}{t}}
\newcommand{\mut}[1]{\coeff{\mu}{#1}{t}}
\newcommand{\setindice}[2]{\left\{#1,\hbox{ }#2\right\}}
\newcommand{\Yun}{Y}
\newcommand{\muun}[1]{\mu_{#1}}

    \newtheorem{theorem}{Theorem}[section]
    \newtheorem{lemma}[theorem]{Lemma}
    \newtheorem{corollary}[theorem]{Corollary}
    \newtheorem{proposition}[theorem]{Proposition}
 \newtheorem{Not}[theorem]{Notation}
    \newtheorem{definition}{Definition}[section]
   \newtheorem{hyp}{Assumption}
    \newtheorem{remark}{Remark}[section]
    
\author{Didier Rullière\footnote{Universit\'e de Lyon, F-69622, Lyon, France; Universit\'e Lyon 1, Laboratoire SAF, EA 2429, Institut de Science Financi\`ere et d'Assurances, 50 Avenue Tony Garnier, F-69007 Lyon, France.~ rulliere@univ-lyon1.fr, diana.dorobantu@adm.univ-lyon1.fr, areski.cousin@univ-lyon1.fr} ,~  Diana Dorobantu\footnotemark[\value{footnote}] ~and Areski Cousin\footnotemark[\value{footnote}]}

\title{ An extension of Davis and Lo's contagion model\thanks{This work has been funded by ANR Research Project ANR-08-BLAN-0314-01. The authors would like to thank the two anonymous referees for useful comments and suggestions. We are grateful to Jean-Paul Laurent for the many helpful discussions on the subject.}}

\begin{document}
\maketitle

\textbf{Abstract} : The present paper provides a multi-period contagion model in the credit risk field. Our model is an extension of Davis and Lo's infectious default model. We consider an economy of $n$ firms which  may default directly or may be infected by other defaulting firms (a domino effect being also possible). The spontaneous default without external influence and the infections are described by not necessarily independent Bernoulli-type random variables. Moreover, several contaminations could be required to infect another firm. In this paper we compute the probability distribution function of the total number of defaults in a dependency context. We also give a simple recursive algorithm to compute this distribution in an exchangeability context. Numerical applications illustrate the impact of  exchangeability among direct defaults and among contaminations, on different indicators calculated from the law of the total number of defaults. \modif{We then examine the calibration of the model on iTraxx data before and during the crisis. The dynamic feature together with the contagion effect seem to have a significant impact on the model performance, especially during the recent distressed period.}\\


\section{Introduction}

The recent financial crisis marked the need for paying more attention to the systemic risk  which can partially be the result of dependence on many factors to a global economic environment.
\modif{
A tractable and common way of modeling dependence among default events is to rely on the conditional independence assumption. Conditionally on the evolution of some business cycle or macroeconomic related factors, defaults are assumed to be independent. However, as shown by some empirical studies such as \cite{Das.Duffie.ea2007}\footnote{The conclusions of this paper have been disputed by \cite{Lando.Nielsen2008} in which the conditional independence assumption has not been totally rejected when tested on the same default database. These discrepancies are explained by an alternative specification of individual default intensities.} or \cite{Azizpour.Giesecke2008}, the latter assumption seems to be rejected when tested on historical default data. An additional source of dependence, namely the chain contagion effect, is observed and requires the construction of contagion models which would be able to explain the \og{}domino effects\fg{}: a defaulting firm causes the default of another firm which infects another one etc. Additional empirical evidences and economic analyses of the default contagion effect can be found in for example \cite{Boissay2006}, \cite{JorionZhang2007} or \cite{JorionZhang2009}}. To capture these effects in a realistic way, multi-period contagion models have to be considered.\\


In this paper we present an extension of \citeauthor{DavisLo}'s model, presented  in
\citeyear{DavisLo} in their paper \og{}Infectious Defaults\fg{} by the authors. They propose a new way of modeling the dependency between defaults
through \og{}infection\fg{}. The original idea applied to  \og{}CBO\fg{} (\textit{Collateralized Bond Obligation}) is that any bond may default either directly or may be infected by any defaulting bond.  Even though in their paper, the authors model contagion effects in a bond portfolio, their study may be  easily applied to any elements  subject to a contagion risk  (for example firms, persons, buildings, etc.). In our paper, the application will focus on credit risk portfolio modeling. In the original model, the direct defaults are described by Bernoulli-independent and identically distributed random variables denoted $X^i$ and the infections are also independent and identically distributed random variables denoted $Y^{ij}$, all Bernoulli distributed ($i \in \N$, $j \in \N$).\\

The present paper provides a model more general than Davis and Lo's. We propose a multi-period model and we further relax the  assumption of independence and identical distribution of the direct defaults and infections. This fact will allow the insertion of possible additional dependencies and to take into account the assets diversity in the portfolio. It will be also possible,  for example, to consider different direct default probabilities  or default dependencies  related to the presence of a common economic factor.

Moreover, compared to Davis and Lo's model in which only  directly defaulting bonds can infect  others, our model allows  taking into account the domino effect which can exist between the bonds, firms etc. Thus in the model presented here, the firms can default because of a chain reaction, phenomena which is often a reason for financial crises.

Finally, the infection way will be generalized here  in order to consider the situations where several infections are necessary to generate a new default : it is possible  for example   to consider the case where only the default of several firms of a sector (and not only one default) can \og{}destabilize\fg{} other firms. These situations can thus limit the contagion phenomena, but also can (together with the contagion probabilities) allow the introduction of possible critical mass of defaults predisposed to make the crisis  more sudden (for example if a large number of infections is necessary to generate other defaults and if the infection probabilities are large).

The model can be used for example for epidemic studies, accidents in chain, or  to introduce in an original manner  a  dependence structure  which could translate the impact of these contaminations.\\

Contagion models were introduced to the credit risk field by \cite{DavisLo} and \cite{JarrowYu}. Since 2001, the financial contagion has become a problem  which drew the attention of many authors. \cite{Yu} proposed an extension of Jarrow and Yu's model. Davis and Lo's infectious model was also studied by many authors who proposed  different models having as starting point Davis and Lo's model (see for example  \cite{Egloff}, \cite{Rosch}, \cite{SakataHisakadoMori}). All along years, the contagion phenomena was modeled using Bernoulli random variables (\cite{DavisLo}), copula functions (\cite{Schonbucher}), interacting particle systems (\cite{Giesecke}), incomplete information models (\cite{FreyRunggaldier2008}) or Markov chains  (see for example \cite{Schonbucher2006}, \cite{GrazianoRogers2006} or \cite{Kraft}).
\modif{
As far as the risk management of synthetic CDO tranches is concerned, Markov chain contagion models have also been investigated by several papers such as \cite{Voort2006}, \cite{Herbertsson.Rootz'en2006}, \cite{Herbertsson2007}, \cite{Frey}, \cite{Frey2008}, \cite{DeKoch.Kraft2007}, \cite{Epple.Morgan.ea2007}, \cite{Lopatin.Misirpashaev2007}, \cite{Arnsdorf.Halperin2007}, \cite{Cont.Minca2008}, \cite{Cont.Deguest.ea2009} among others. The hedging issue for CDO tranches is also addressed by \cite{Laurent} and \cite{Cousin_Jeanblanc_Laurent2009} in the class of Markovian contagion models. As a first step among possible applications of the Davis and Lo's multi-period model, we illustrate the model tractability in terms of the pricing of synthetic CDO tranches. Since the computation of CDO tranche spreads involves the computation of expected tranche losses at several time horizons, such a study could not have been done using the original static version of the model. We then discuss the model performance in terms of fitting liquid CDO tranche spreads.}\\


In the paper \textit{Infectious Defaults}, Mark Davis and Violet
Lo present a model where the bonds  may default either directly or may be infected by any defaulting bond
\citep[see][]{DavisLo}. The authors consider $n$ bonds, $n \in \N^*$. Let $\Omega$  be the set of possible indices of these bonds, $\Omega=\{ 1, .., n \}$. For a bond $i$, $i \in \Omega$, the random variable $X^i$ is equal to $1$ if the bond defaults directly, $0$ otherwise. Even if $X^i=0$,
bond $i$ can be defaulted by infection (contagion). A default infection takes place  if a Bernoulli random variable $\C^i=1$. Thus the bond $i$ is defaulted  (directly  or by infection) if  $Z^i=1$, with
$$Z^i=X^i + (1-X^i) \C^i.$$

In  \citeauthor{DavisLo}'s  paper, the infection is modeled in the following way :
$$\C^i=1-\prod_{j\in \Omega, j \neq i} \left(1-X^j Y^{ji} \right),$$
where $Y^{ji}$, $i,j \in \Omega$ are Bernoulli random variables. Consequently, at the end of the period, a bond $i$ is infected if at least one  $X^j Y^{ji}=1$, $j \in \Omega$, $j \neq
i$, i.e. \textit{at least another} bond $j$ defaults directly
($X^j=1$) and infects the first bond  ($Y^{ji}=1$).

\modif{In the following, $\bino nk$ will denote the binomial coefficient}\modif{ ($0\le k \le n$) : $\bino nk=\frac{n!}{k!(n-k)!}$}.

Davis and Lo stated and proved the following theorem :
\begin{theorem}[\citeauthor{DavisLo}'s result]
\begin{eqnarray*}
\Proba{\sum_{i \in \Omega} Z^i=k} &=& \bino nk \alpha_{nk}^{pq},\\
\hbox{where } \alpha_{nk}^{pq} &=& p^k (1-p)^{n-k} (1-q)^{k (n-k)} +
\sum_{i=1}^{k-1} \bino ki p^i (1-p)^{n-i} (1-(1-q)^i)^{k-i} (1-q)^{i
(n-k)}.
\end{eqnarray*}
\end{theorem}

In this paper, we  propose to relax the following assumptions of the original model :

\begin{itemize}
  \item
  A1. The random variables $\set{X^i, ~i\in \Omega}$ are independent and identically distributed and  $\forall ~i \in \Omega$, $\Proba{X^i=1}=p$, $p\in]0, 1[$,
  \item
   A2. The random variables $\set{Y^{ij}, ~(i, j) \in \Omega^2}$ are independent and identically distributed  and $\forall ~i,j \in \Omega$, $j \neq i$, $\Proba{Y^{ij}=1}=q$, $q\in]0, 1[$,
  \item
   A3. The infection $\C^i$ occurs if at least one  $X^j Y^{ji}=1$, $j \in \Omega$, $j \neq i$.
 \item
   A4. An infected bond can not infect another bond  (one period model).\\
\end{itemize}

The outline of the present paper is as follows : we present the  model and we fix the assumptions under which we  will state our theorems  (Section \ref{modele}). We present some life insurance tools used for the characterization of the defaults' number law  (Section \ref{outils}). In the following Section \ref{1_per} we give the defaults' number law in a one period model. Our result allows  finding, under the random variables independence assumption,  Davis and Lo's result. In  Section \ref{T_per}, we generalize the model of the previous section and we present a multi-period model. Finally  we give some numerical applications (see Section \ref{applic}) \modif{including an illustration of parameters effects on the loss distribution and a discussion on the ability of the model to fit iTraxx CDO tranche quotes.}

\section{The Model}
\label{modele}

We consider an economy of $n$ firms ($n\in\N^*$). Let $\Omega$ be the set of possible indices of these firms : $\Omega=\{1, ..., n\}$. We observe the firms along a given  time interval  divided into $T$ periods.   When the firm $i$ ($i\in\Omega$) defaults directly during the period $t$ ($t\in\{1, ..., T\}$), then the random variable $X^i_t$ is equal to $1$, otherwise $X^i_t=0$. Even if $X^i_t=0$, the firm $i$ can be defaulted during the period $t$ : it may be infected by another defaulting firm. We use a binary random variable $\C^i_{t}$ to denote this infection. This variable is equal to $1$ when the firm $i$ does not default directly but is infected by other firms during the period $t$, $\C^i_{t}=0$ otherwise. Let $\{Z_t^i, ~i\in\Omega\}$ be a sequence of discrete random variables with possible values $\{0, ~1\}$. For any $i\in\Omega$, $Z^i_t=1$ if the firm $i$ is defaulted (directly or   by infection)  at the end of period $t$,  $Z^i_t=0$ otherwise. In our model, the random variables $\{Z_t^i, ~t=1, ..., T\}$ are obtained by the following recursive relation :
\begin{align*}
Z^i_t&=Z^i_{t-1}+(1-Z^i_{t-1})[X_t^i+(1-X_t^i) \C^i_{t}], ~~ 2\leq t\leq T, ~i\in\Omega,\\
Z^i_1&=X^i_1+(1-X_1^i) \C^i_{1}.
\end{align*}

Hence $Z^i_t=1$ if the firm $i$ has been declared in default at the end of period $t-1$ ($Z^i_{t-1}=1$) or if it defaults directly  ($X_t^i=1$) or by contagion  ($\C^i_{t}=1$) during the period  $t$.

We introduce the following notations :
\begin{Not}\label{not1}
For every $t\in\set{1, \dots T}$, we denote by :
\begin{itemize}
\item $\Theta_t$ the set of the firms declared in default up to period $t$ : $\Theta_t=\set{i\in\Omega, \, Z_t^i=1}$,
\item $\Gamma_t$ the set of the firms which did not default in the previous $t$ periods : $\Gamma_t=\set{i\in\Omega, \, Z_t^i=0}=\Omega-\Theta_t$,
\item $N_t^D$ the number of spontaneous defaults without external influence occurred \modif{at time} $t$ : \\$N_t^D = \sum_{i \in \Gamma_{t-1}} X_t^i$,
\item $N_t$ the \modif{cumulated number of  defaults occurred up to time $t$} : $N_t=\sum_{i \in \Omega} Z_t^i$.
\end{itemize}
\end{Not}

For every $t\in\set{1, \dots T}$, the random variable $\Yt^{ji}$, $((j, i)\in\Omega^2)$ with possible values $\{0, ~1\}$ represents the infection during the period $t$ of the firm $i$  by a defaulting firm $j$. Consequently the variables $\C^i_{t}$ ($i\in\Omega$) have the form
 $$\C^i_t=f\left(\sum_{j\in F_t} \Yt^{ji}\right),$$
 where $f: \{0,..,n\}\rightarrow\{0,1\}$ and $F_t$ represents the set of the defaulting firms susceptible to infect the  firm $i$. Here we suppose that $card(F_t)=g(N_{t-1}, ~N_t^D),$ where $g : \N^2\rightarrow\{ 0, .., n \}$. For example, in the original model $F_1=N_1^D$ (thus only the firms which default directly  may infect  others) and $f(i)=\Indic{i \ge 1}$ i.e. at least a contamination causes the default.
In our model, we can imagine the case where the firm $i$ defaults only if  it is infected by a given number of firms, for example $f(i)=\Indic{i \ge 2}$ (i.e. at least two infections are necessary to generate a new default). Moreover, the firms which can infect the firm $i$ during a given period $t$,  are  those which were directly defaulted during the same period  $t$, or those  declared in default  (directly or by infection) before this  period.

The aim  of this paper is to study the law of $N_t$ under the following assumptions  :
\begin{hyp}[Temporal independence of direct defaults]
\label{h1}
$ $\\
The random vectors $\vec{X}_t=(X^1_t, ...,X^n_t)$, $t\in\set{1, \dots, T}$, are mutually independent, but a dependency exists between  the  components of each vector.
 \end{hyp}
 \begin{hyp}[Temporal independence of exchangeable infections]
\label{h2}
$ $\\
The random vectors  $\vec{Y}_t=(\Yt^{11}, \Yt^{12},...,\Yt^{nn})$, $t\in\set{1, \dots, T}$, are mutually independent and for any $t$, the random variables   $\set{\Yt^{ji}, ~(j, i)\in \Omega^2}$ are exchangeable, independent of $\set{X^i_t, ~t=1,\dots,T, ~i\in\Omega}$.
\end{hyp}

Under these assumptions, we suppose that all joint distributions of $(X_t^1, \dots, X_t^n)$, $t\in\set{1, \dots, T}$ and of $(\Yt^{\sigma(1)}, \dots, \Yt^{\sigma(n^2)})$ are known, where  $\Yt^{\sigma()}$ is a permutation of the set  $\Yt^{ij}$, $(i, j)\in\Omega^2$.

\modif{We will sometimes use the following assumption which is a particular case of Assumption \ref{h1} : }
\begin{hyp}[Temporal independence of exchangeable direct defaults]
\label{h4}
$ $\\
The random vectors $\vec{X}_t=(X^1_t, ...,X^n_t)$, $t\in\set{1, \dots, T}$, are mutually independent, but the components of each vector are exchangeable random variables.
 \end{hyp}
\modif{In the particular case of infinite exchangeability assumption}, thanks to De Finetti's Theorem, this assumption is perfectly adapted to the situation where the direct defaults are dependent because of a \og{}hidden\fg{} common factor  (for example the state of the economy).

\modif{At last we will sometimes consider the particular case of} \modif{independent and identically distributed (i.i.d.)} \modif{  direct defaults and i.i.d. contaminations :}
\begin{hyp}[i.i.d. direct defaults and i.i.d. contaminations]
\label{h5}
$ $\\
The random variables $\set{X^i_t, ~i\in\Omega, ~t\in\set{1, \dots, T}}$  are independent, all Bernoulli distributed with parameter $p$ and $\set{\Yt^{ij}, ~(i, j)\in\Omega^2, ~t\in\set{1, \dots, T}}$ are also  independent random variables, all Bernoulli distributed with parameter $q$.
 \end{hyp}

Compared to Davis and Lo's infectious model, we propose a multi-period model where  $\set{X^i_t, ~i\in\Omega}$ and $\set{\Yt^{ij}, ~(i, j)\in \Omega^2}$ are no more independent and  identically distributed random variables. Moreover, we do not consider just the particular case where  $f(i)=\Indic{i \ge 1}$ and the firms which can infect are not necessary those which default directly during the same period.  Thus, we take into account the \og{}domino effect\fg{}, i.e. the firm $i$ defaults, it infects the firm $j$ which infects the firm $k$ etc.

\modif{To generalize the mode of infection, one can for example consider, $f(i)=\Indic{i \ge 2}$ instead of $f(i)=\Indic{i \ge 1}$. In other words, two contagions are required to cause an indirect default. This feature of our model allows to weaken the effect of contagion. Such a possibility can be useful for large portfolios.} Indeed, it is important to generalize the mode of infection. Let us imagine that the number of firms  increases. Then, if all the parameters of the model remain identical, too many infections could occur.  Other recent generalizations were proposed and they introduce different solutions to generalize the mode of infection in order to better represent the credit risk markets  reality
\citep[see for example][]{SakataHisakadoMori}. Furthermore, a generalization may allow to better model the critical thresholds up to which the infections start to have a real effect. \modif{Different modes of contagion will be compared in numerical applications (see Subsection \ref{Sec:Effect_of_parameters}).}

\section{Results}

\modif{The aim of this section is to present the main results of the paper together with some tools used to prove them.}
\subsection{Life Insurance Tools}
\label{outils}

In this section we present some useful tools to characterize the defaults number distribution.

We introduce the coefficients of order  $k$ for the set $\set{X_t^i, ~i\in\Gamma}$ where $\Gamma\subset\Omega$, $card(\Gamma)\geq k$, $k\in\N$  :
\begin{definition}
Let $\Gamma\subset\Omega$. For all $t\in\set{1, \dots, T}$, the coefficient of order $k$ ($k\leq card(\Gamma)$) for the set
 $\set{X_t^i, ~i\in\Gamma}$, denoted $\mut{k}(\Gamma)$, is defined as
\begin{align*}
\mut{k}(\Gamma)&=\frac{1}{\bino{card(\Gamma)}{k}} \displaystyle\sum_{\substack{j_1<j_2<..<j_k \\ j_1, \dots, j_k\in\Gamma}} \Proba{
X_t^{j_1}=1 \cap ... \cap X_t^{j_k}=1}, ~~1\leq k\leq card(\Gamma),\\
\mut{0}(\Gamma)&=1 \hbox{~(including if~} \Gamma=\emptyset).
\end{align*}
The  $\displaystyle\sum_{\substack{j_1<j_2<..<j_k \\ j_1, \dots, j_k\in\Gamma}}$ symbol means to sum all  $\bino{card(\Gamma)}{k}$ possible choices of $k$ different elements taken among the elements of the set $\Gamma$.
\end{definition}

\begin{remark}
\label{rem1}
If the random variables $\set{X_t^i, ~i\in\Omega}$ are exchangeable, then for all  $\Gamma\subset\Omega$  we have
$$\mut{k}(\Gamma)=\mut{k}= \Proba{X_t^{1}=1 \cap ... \cap X_t^{k}=1},  ~1\leq k\leq card(\Gamma).$$
In the particular case where  $\set{X_t^i, ~i\in\Omega}$ are independent random variables, all Bernoulli distributed with parameter $p$, then $\mut{k}=p^k$.
\end{remark}

A classical problem in the actuarial field is to compute the number of survivors after a given period, among a group of different persons with independent and identically distributed  survival rates. We will see that these tools can considerably simplify the studied credit risk models.

\begin{lemma}[Waring formula]
\label{Waring}
Let $X_t^1,...,X_t^n$ be $n$ dependent Bernoulli random variables and let $\Gamma$ be a subset of $\set{1, \dots, n}$ such that $card(\Gamma)=m\leq n$. Then
$$\Proba{\sum_{i\in\Gamma} X_t^i=k}=\Indic{k\leq m} \bino mk \sum_{j=0}^{m-k} \bino{m-k}{j} (-1)^j \mut{j+k}(\Gamma).$$
\end{lemma}

This result was proved  by  \cite{Feller} (chapter IV.3, page 106). An extension of this formula is  Schuette-Nesbitt Formula often used in actuarial science (see \cite{Gerber}, chapter 8.6, page 89).
\begin{remark}
\label{rem2}
Let $X_t^1,...,X_t^n$ be $n$ exchangeable Bernoulli random variables and let $M_X=\set{i : X_t^i=1}$. Then for all $E_k\subset\set{1, \dots, n}$ such that $card(E_k)=k$ we have the following equalities :
$$\Proba{\sum_{i=1}^n X_t^i=k}= \bino nk \Proba{X_t^1=\dots=X_t^k=1, ~X_t^{k+1}=\dots=X_t^n=0}= \bino nk \Proba{M_X=E_k}.$$
\end{remark}

Since the variables $\set{\Yt^{ij}, ~(i, j)\in\Omega^2}$ are exchangeable, then conditionally to  $\set{N_{t-1}, N_t^D}$, the variables $\set{\C_t^i, ~i\in\Omega}$ are also exchangeable. We introduce the coefficient of order $k$ for these two sets of random variables.
\begin{definition}
For all $\Gamma\subset\Omega^2$ and for all $t\in\set{1, \dots T}$, the coefficient of order $k$ ($k\leq card(\Gamma)$)  for the set
 $\set{\Yt^{ij}, ~(i, j)\in\Gamma}$, denoted $\lambdat{k}$, is defined as
\begin{align*}
\lambdat{k}=& \Proba{\Yt^{\sigma(1)}=1 \cap ... \cap \Yt^{\sigma(k)}=1}, ~k\geq 1,\\
\lambdat{0}=&1.
\end{align*}
In  the same way we define
\begin{align*}
\xi_{k,t}(g(u,l))=&\Proba{\C_t^{1}=1 \cap ... \cap \C_t^{k}=1\sachant N_{t-1}=u, N_t^D=l}, ~k\geq 1,\\
\xi_{0,t}(g(u,l))=&1 \hbox{~(including if~} g(u,l)=0).
\end{align*}
\end{definition}

Every variable $\C_t^{i}$ is a function of $\set{\Yt^{ji}, ~j\in F_t}$ which are exchangeable random variables.  The law of $(\C^1_t, \dots, \C^n_t)$ depends only on $card(F_t)$ and on the law of $(\Yt^{\sigma(1)}, \dots, \Yt^{\sigma(card(F_t))})$. We can thus  establish a  relationship between the coefficients  $\lambdat{k}$ and $\xi_{k,t}(z)$, $z\leq n$.

\begin{proposition}[Joint law of $(\C_t^{1}\dots \C_t^{k})$]
\label{prop1}
$ $\\
The joint law of $(\C_t^{1}\dots \C_t^{k})$ is given by $$\xi_{k,t}(z)=\Proba{f(\sum_{j\in F_t} \Yt^{j1})=1 \cap ... \cap f(\sum_{j\in F_t} \Yt^{jk})=1\sachant card(F_t)=z}, ~0\leq k\leq n, ~0\leq z\leq n-k.$$
\begin{enumerate}
\item
In the particular case where $f(i)=\Indic{i \ge 1}$,
 we have
$$\xi_{k,t}(z)=\sum_{i=0}^{k} \bino ki \sum_{\alpha=0}^{zi} \bino {zi}{\alpha}(-1)^{i+\alpha}\lambdat{\alpha}.$$
\item
For any function $f : \{0, \dots, n\}\rightarrow \{0, ~1\}$, we have
$$\xi_{k,t}(z)=\sum_{\gamma=0}^{zk}\eta_{k, z}(\gamma)\sum_{j=0}^{zk-\gamma} \bino{zk-\gamma}{j}(-1)^j\lambdat{j+\gamma},$$
where $\eta_{k, z}(\gamma)=\displaystyle\sum_{\substack{\gamma_1\in \set{0, \dots, z}\\\gamma_1\leq\gamma}}f(\gamma_1) \bino{z}{\gamma_1}\eta_{k-1, z}(\gamma-\gamma_1),$
$\eta_{1, z}(x)=\Indic{x\leq z}f(x) \bino{z}{x}$ and $\eta_{0, z}(x)=\Indic{x=0}.$
\end{enumerate}
\end{proposition}
This proposition is proved in  Appendix.
\begin{remark}
\label{indepxi}
If  $\set{Y_t^{ij}, ~(i, j)\in\Omega^2, ~t\in\set{1,\dots,T}}$ are independent random variables,  all Bernoulli distributed with parameter  $q$, then $\lambdat{k}=q^{k}$ and
\begin{itemize}
\item
in the particular case where $f(i)=\Indic{i \ge 1}$,   $\xi_{k,t}(z)=(1-(1-q)^z)^k$,
\item
for any function $f : \{0, \dots, n\}\rightarrow \{0, ~1\}$,
$\xi_{k,t}(z)=\displaystyle\sum_{\gamma=0}^{zk}\eta_{k, z}(\gamma) q^{\gamma}(1-q)^{zk-\gamma}.$
\end{itemize}
\end{remark}

\subsection{One Period Model}
\label{1_per}

The aim of this section is to characterize the defaults' number distribution in the case where   $T=1$.
\modif{For the sake of clarity, all subscripts $1$ are omitted, so that $N=N_1$, $X^i=X^i_1$, $Z^i=Z^i_1$...}

It is a one period model which generalizes the \citeauthor{DavisLo}'s results when the direct defaults and the infections are not  necessary  independent. In the particular case where $T=1$, the random variables $\set{Z^i, ~i\in\Omega}$ are obtained as in the  \citeauthor{DavisLo}'s model :
$$Z^i=X^i+(1-X^i) \C^i, ~i\in\Omega,$$
where  $\C^i=f(\displaystyle\sum_{j\in F_t} \Yun^{ji})$, $card(F_t)=\tilde{g}(N^D)$ (with $\tilde{g}(x)=g(0, x)$). Let us recall that in the original model, the authors consider only the particular case where $f(i)=\Indic{i\geq 1}$.

Under Assumption  \ref{h1}, the random variables $\set{X^i, ~i\in\Omega}$ are no more independent. We suppose that all the joint laws of $(X^1, \dots, X^n)$ and $(\Yun^{\sigma(1)}, \dots, \Yun^{\sigma(n^2)})$ are known.

One of the main results of this paper is the following theorem which gives the defaults' number law.
\begin{theorem}[Defaults' number law when  the defaults are not i.i.d. - One period model]
\label{th1}
$ $
\\Under Assumptions \ref{h1} and \ref{h2}, if $T=1$, then the defaults' number law is given by :
\begin{eqnarray}
\nonumber
\Proba{N=r}= \bino nr \sum_{k=0}^{r} \bino rk \sum_{\alpha=0}^{n-r}   \bino{n-r}{\alpha} (-1)^{\alpha}\xi_{\alpha+r-k, 1}(\tilde{g}(k))\sum_{j=0}^{n-k} \bino{n-k}{j} (-1)^j\muun{j+k}(\Omega).
\end{eqnarray}
\end{theorem}
\Proof
Using the law of alternatives, we get
\begin{eqnarray}
\nonumber
\Proba{N=r}=\sum_{k=0}^{r}\Proba{N=r\sachant N^D=k}\Proba{ N^D=k}.
\end{eqnarray}
On the one hand, by Lemma \ref{Waring}, $\Proba{ N^D=k}= \bino nk \displaystyle\sum_{j=0}^{n-k} \bino{n-k}{j} (-1)^j\muun{j+k}(\Omega)$.
\\
On the other hand \begin{eqnarray}
\nonumber
\Proba{N=r\sachant N^D=k}=\Proba{\sum_{i=1}^{n}X^i+(1-X^i)\C^i =r\sachant N^D=k}=\Proba{\sum_{i\in A}\C^i =r-k\sachant N^D=k}
\end{eqnarray}
 where $A=\set{i\in\Omega : X^i=0}.$
\\
Since $card(A)=n-k$,  Lemma \ref{Waring} gives $\Proba{N=r\sachant N^D=k}= \bino{n-k}{r-k} \sum_{\alpha=0}^{n-r} \bino{n-r}{\alpha} (-1)^{\alpha}\xi_{\alpha+r-k, 1}(\tilde{g}(k)).$ The equality $ \bino nk \bino{n-k}{r-k} = \bino nr \bino rk$ concludes the proof.
\EndProof

We easily deduce the law of $N$ when  $\set{X^i, ~i\in\Omega}$ is a set of exchangeable variables.

\begin{corollary}[Defaults' number law when  the defaults are exchangeable - One period model]
\label{cor1}
$ $\\
\modif{Under Assumptions~\ref{h2} and~\ref{h4}}, \modif{assume that the variables $\set{X^i, ~i\in\Omega}$ are exchangeable. Then the defaults' number law is given by :}
\begin{eqnarray}
\nonumber
\Proba{N=r}= \bino nr \sum_{k=0}^{r} \bino rk \sum_{\alpha=0}^{n-r} \bino{n-r}{\alpha} (-1)^{\alpha}\xi_{\alpha+r-k, 1}(\tilde{g}(k))\sum_{j=0}^{n-k} \bino{n-k}{j} (-1)^j\muun{j+k},
\end{eqnarray}
where the coefficients $\muun{j}$ are introduced in Remark \ref{rem1}.
\end{corollary}

As a consequence of Theorem \ref{th1}, Remarks \ref{rem1} and  \ref{indepxi}, is the following result which gives the defaults' number law in the particular case where  $\set{X^i, ~i\in\Omega}$ are independent  random variables, all Bernoulli distributed with parameter $p$ and $\set{\Yun^{ij}, ~(i, j)\in\Omega^2}$ are also  independent random variables, all Bernoulli distributed with parameter $q$.

\begin{corollary}[Defaults' number law when  the defaults are independent - One period model]
\label{cor2}
$ $\\
Under Assumption~\ref{h5}, \modif{assume that}  $\set{X^i, ~i\in\Omega}$ are independent  random variables, all Bernoulli distributed with parameter $p$ and $\set{\Yun^{ij}, ~(i, j)\in\Omega^2}$ are also  independent random variables, all Bernoulli distributed with parameter $q$. Then the defaults' number law is given by
\begin{eqnarray}
\nonumber
\Proba{N=r}= \bino nr \sum_{k=0}^{r} \bino rk p^k(1-p)^{n-k}\sum_{\alpha=0}^{n-r}    \bino{n-r}{\alpha} (-1)^{\alpha}\xi_{\alpha+r-k}(\tilde{g}(k)),
\end{eqnarray}
where
$\xi_{u}(z)=\displaystyle\sum_{\gamma=0}^{zu}\eta_{u, z}(\gamma)q^{\gamma}(1-q)^{zu-\gamma}.$
\end{corollary}

Under the assumptions of Corollary \ref{cor2},  we also assume that at least one infection causes a new default  (i.e. $f(i)=\Indic{i \ge 1}$). Then, using Remark \ref{indepxi} and taking $\tilde{g}$ the identity function we find the Davis and Lo's result.

\subsection{Multi-period Model}
\label{T_per}

Generalizing the model presented in the previous section, we consider now  that the time interval is divided into several periods.  In this model the default indicator is described by the relation :

\begin{align}
\label{Eq:default_indicators_multi_period_1}
Z^i_t&=Z^i_{t-1}+(1-Z^i_{t-1})[X_t^i+(1-X_t^i)\C^i_{t}], ~~ 2\leq t\leq T, ~i\in\Omega,\\
\label{Eq:default_indicators_multi_period_2}
Z^i_1&=X^i_1+(1-X_1^i)\C^i_{1}.
\end{align}


We are interested in computing the law of the defaults' number at the end of every period $t$, \\$t\in\set{1, \dots, T}$, $T>1$. Another main result of this paper is the following theorem :
\begin{theorem}[Defaults' number law when  the defaults are not i.i.d. - Multi-period model]
\label{th2}
$ $\\
Under Assumptions \ref{h1} and \ref{h2}, the defaults' number law is given by :
\begin{eqnarray}
\nonumber
\Proba{N_t=r}=\sum_{\substack{\theta_t\subset\Omega\\ card(\theta_t)=r}}\Proba{\Theta_t=\theta_t},
\end{eqnarray}
where $\Theta_t$ is the set of firms declared in default up to  $t$ (see Notations~\ref{not1}) and where
\begin{align*}
\Proba{\Theta_t=\theta_t}&=\displaystyle\sum_{u=0}^r\sum_{\substack{\theta_{t-1}\subset\theta_t\\ card(\theta_{t-1})=u}}\Proba{\Theta_t=\theta_t\sachant\Theta_{t-1}=\theta_{t-1}}\Proba{\Theta_{t-1}=\theta_{t-1}},\\
\Proba{\Theta_t=\theta_t\sachant\Theta_{t-1}=\theta_{t-1}}&=\displaystyle\sum_{m=0}^{r-u}\sum_{\substack{M_t\subset \theta_t-\theta_{t-1}\\card(M_t)=m}}\rho(M_t, \Omega-\theta_{t-1}-M_t)\sum_{j=0}^{n-r} \bino {n-r}{j} (-1)^j\xi_{j+r-u-m, t}(u, m),\\
\rho(A, ~B)&=\Proba{\forall ~i\in A ~~X_t^i=1 \hbox{~et~} \forall ~i\in B ~~X_t^i=0} ~~~\forall ~A, B\subset\Gamma_{t-1},
\end{align*}
with $card(\theta_t)=r$ and $card(\theta_{t-1})=u$, $u\leq r$.
\end{theorem}

The first term $\Proba{\Theta_1=\theta_1}$ may be easily computed using the recursive formula for $\Proba{\Theta_1=\theta_1\sachant\Theta_0=\emptyset}.$

\Proof
We split $$\Proba{N_t=r}=\displaystyle\sum_{\substack{\theta_t\subset\Omega\\ card(\theta_t)=r}}\Proba{\Theta_t=\theta_t}.$$
By the law of alternatives
$$\Proba{\Theta_t=\theta_t}=\displaystyle\sum_{u=0}^r\sum_{\substack{\theta_{t-1}\subset\theta_t\\ card(\theta_{t-1})=u}}\Proba{\Theta_t=\theta_t\sachant\Theta_{t-1}=\theta_{t-1}}\Proba{\Theta_{t-1}=\theta_{t-1}}.$$
It thus remains to compute the term $\Proba{\Theta_t=\theta_t\sachant\Theta_{t-1}=\theta_{t-1}}$ when $\theta_{t-1}\subset\theta_t\subset\Omega$, $card(\theta_t)=r$ and $card(\theta_{t-1})=u$, $u\leq r$.
$$\Proba{\Theta_t=\theta_t\sachant\Theta_{t-1}=\theta_{t-1}}=\Proba{\forall ~i\in\theta_t-\theta_{t-1}, ~X^i_t+(1-X^i_t)\C^i_t=1 \hbox{~et~} \forall ~i\in\Omega-\theta_t, ~X^i_t=\C^i_t=0\sachant\Theta_{t-1}=\theta_{t-1}}.$$ 

Using Remark \ref{rem2} and then Lemma \ref{Waring}, we obtain that $\Proba{\Theta_t=\theta_t\sachant\Theta_{t-1}=\theta_{t-1}}$ is equal to
\begin{align*}
&\displaystyle\sum_{m=0}^{r-u}\sum_{\substack{M_t\subset \theta_t-\theta_{t-1}\\card(M_t)=m}}\rho(M_t, \Omega-\theta_{t-1}-M_t)\frac{\Proba{\sum_{i\in\Omega-\theta_{t-1}-M_t}\C^i_t=r-u-m\sachant N_{t-1}=u, N_t^D=m}}{\bino{n-u-m}{r-u-m}}\\
=&\sum_{m=0}^{r-u}\sum_{M_t\subset\theta_t-\theta_{t-1}}\rho(M_t, ~\Omega-\theta_{t-1}-M_t)\sum_{j=0}^{n-r} \bino {n-r}{j}(-1)^j\xi_{j+r-u-m, t}(g(u, m)).
\end{align*}
\EndProof

If the direct defaults are not i.i.d., the law of $N_t$ is given by a recursive relation (see Theorem \ref{th2}). However, this relation can  induce an important load of calculation in  numerical applications. That is why in the numerical applications we will consider the case where the random variables $\set{X^i_t, ~i\in\Omega, ~t\in\set{1, \dots, T}}$ satisfy the \modif{Assumption~\ref{h4} and are exchangeable}.

The following result is a particular case of Theorem \ref{th2}.

\begin{theorem}[Defaults' number law when  the defaults are exchangeable - Multi-period model]
\label{th3}
$ $\\
Under Assumptions  \ref{h2} and \ref{h4}, the defaults' number law is given by
\begin{eqnarray}
\nonumber
\Proba{N_t=r}=\sum_{k=0}^r\Proba{N_{t-1}=k} \bino{n-k}{r-k}\sum_{\gamma=0}^{r-k} \bino{r-k}{\gamma}\sum_{\alpha=0}^{n-k-\gamma} \bino{n-k-\gamma}{\alpha}\mut{\gamma+\alpha}\sum_{j=0}^{n-r} \bino{n-r}{j}(-1)^{j+\alpha}\xi_{j+r-k-\gamma, t}(g(k, \gamma)).
\end{eqnarray}
\end{theorem}

The proof of this theorem rests on the following lemma shown in Appendix :
\begin{lemma}
\label{lemme2}
Under Assumptions  \ref{h2} and \ref{h4}, for any $\theta_t$ and $\theta_{t-1}$ such that $\theta_{t-1}\subset\theta_t\subset\Omega$, $card(\theta_t)=r$, $card(\theta_{t-1})=k$, $k\leq r$, the following relation is true :
$$\Proba{N_t=r\sachant N_{t-1}=k}= \bino{n-k}{r-k}\Proba{\Theta_t=\theta_t\sachant \Theta_{t-1}=\theta_{t-1}}.$$
\end{lemma}
\preuve \textbf{of Theorem \ref{th3}}
Using the law of alternatives and then Lemma \ref{lemme2}, we obtain :
\begin{align*}
\Proba{N_t=r}&=\sum_{k=0}^r\Proba{N_t=r\sachant N_{t-1}=k }\Proba{N_{t-1}=k }\\
&=\sum_{k=0}^r\Proba{N_{t-1}=k } \bino{n-k}{r-k}\Proba{\Theta_t=\theta_t\sachant \Theta_{t-1}=\theta_{t-1}},
\end{align*}
for any $\theta_t$ and $\theta_{t-1}$ such that $\theta_{t-1}\subset\theta_t\subset\Omega$, $card(\theta_t)=r$, $card(\theta_{t-1})=k$, $k\leq r$.

By Theorem \ref{th2},
$$\Proba{\Theta_t=\theta_t\sachant\Theta_{t-1}=\theta_{t-1}}=\sum_{\gamma=0}^{r-k}\displaystyle\sum_{\substack{M_t\subset \theta_t-\theta_{t-1}\\ card(M_t)=\gamma}}\rho(M_t, \Omega-\theta_{t-1}-M_t)\sum_{j=0}^{n-r} \bino{n-r}{j}(-1)^j\xi_{j+r-k-\gamma, t}(g(k, \gamma)).$$
However, under Assumptions \ref{h2} and \ref{h4},
$\rho(M_t, \Omega-\theta_{t-1}-M_t)=\sum_{\alpha=0}^{n-k-\gamma} \bino{n-k-\gamma}{\alpha} (-1)^{\alpha}\mut{\gamma+\alpha}.$
We remark that  $\displaystyle\sum_{\substack{M_t\subset \theta_t-\theta_{t-1}\\ card(M_t)=\gamma}}1= \bino{r-k}{\gamma}$ which concludes the proof.
\EndProof

We easily deduce the law of  $N_t$ when all the variables are mutually independent.
\begin{corollary}[Defaults' number law when the defaults are independent - Multi-period model]
\label{cor10}
$ $\\
\modif{Under Assumption~\ref{h5}}, \modif{assume that} \modif{both $\set{X^i_t}$ and $\set{\Yt^{ij}}$ are i.i.d. Bernoulli random variables of respective parameters $p$ and $q$.} The defaults' number law is given by
\begin{eqnarray}
\nonumber
\Proba{N_t=r}=\sum_{k=0}^r\Proba{N_{t-1}=k} \bino{n-k}{r-k}\sum_{\gamma=0}^{r-k} \bino{r-k}{\gamma}p^{\gamma}(1-p)^{n-k-\gamma}\sum_{j=0}^{n-r} \bino{n-r}{j}(-1)^{j+\alpha}\xi_{j+r-k-\gamma, t}(g(k, \gamma)),
\end{eqnarray}
where
$\xi_{u,t}(z)=\displaystyle\sum_{j=0}^{zu}\eta_{u, z}(j) q^{j}(1-q)^{zu-j}.$

Moreover, if $f(i)=\Indic{i \ge 1}$, then
\begin{eqnarray}
\nonumber
\Proba{N_t=r}=\sum_{k=0}^r\Proba{N_{t-1}=k} \bino{n-k}{r-k}\sum_{\gamma=0}^{r-k} \bino{r-k}{\gamma}p^{\gamma}(1-p)^{n-k-\gamma} (1-(1-q)^{g(k, \gamma)})^{r-k-\gamma}(1-q)^{g(k, \gamma)(n-r)}.
\end{eqnarray}
\end{corollary}

Using the last equation of Corollary  \ref{cor10} for $t=1$, $g(k,\gamma)=\gamma$, and $N_0=0$ almost surely, we find the \citeauthor{DavisLo}'s result.

\section{Numerical applications} \label{applic}

\modif{In this section, we provide some numerical applications of the previous extension of Davis and Lo's model.\\

We first investigate the effect} of exchangeability, among direct defaults and among infections, on the total defaults' number evolution. This kind of impact is also dealt with by~\cite{Denuit1} and~\cite{Denuit} in a life insurance framework, and it appears that some quantitative measures could be notably affected by dependencies among the considered random variables. Here, we will  be concerned about the impact of such dependencies on the first moments of total defaults' number $N_t$ and on the survival function of $N_t$ at some points.\\

\modif{We then study the ability of the model to price synthetic CDO tranches. Interestingly, the first application proposed in the original paper by Davis and Lo focuses on the rating of Collateralized Bond Obligations. However, to the best of our knowledge, no calibration procedure of the Davis and Lo's contagion model has been tested so far on liquid tranche quotes. In this framework, direct defaults are considered to be Bernoulli mixtures and contagion events are assumed to be independent. Using this specification, we compare the calibration performance of the model on several years of iTraxx data before and during the crisis.}

\subsection{General settings}

\subsubsection{Algorithm in the exchangeable case}
\label{Seq:Algo}

\modif{Let us}   consider a framework where both direct defaults and infections are described by the following two sets of exchangeable random variables $\setindice{X^i_t}{i \in \Omega}$ and $\setindice{\Yt^{ij}}{(i,j) \in \Omega^2}$. So that Assumptions~\ref{h2} and~\ref{h4} hold. Furthermore,   suppose that the  random vector laws do not change over time. As a consequence, in order to lighten further notations, we  omit the $t$ subscript in coefficients $\xi_{k,t}(z)$, $\mut{k}$ and $\lambdat{k}$, which \modif{become} $\xi_k(z)$, $\mu_k$ and $\lambda_k$. The presented algorithms could be easily adapted in the case where those quantities vary over time.

From Theorem~\ref{th3}, we have seen that the law characterization of total defaults' number was requiring the computation of the set of coefficients $\setindice{\xi_{k}(z)}{k \in \set{0,\dots,n}, z \in \set{0,\dots,n-k}}$. In the general case where $f$ is not necessarily the indicator function $f(x)=\Indic{x \ge 1}$, given initialization values that are provided in section~\ref{outils}, and given the Proposition~\ref{prop1}, this set of coefficients can be computed by the following algorithm:
\begin{proposition}[Algorithm for $\setindice{\xi_k(z)}{k,z \in \N}$]\label{algo1} The coefficients $\setindice{\xi_k(z)}{k,z \in \N, k+z\le n}$ can be computed by:
\begin{itemize}
\item Initialization: for $z$ varying from $0$ to $n$, $\eta_{0,z}(0) = 1$, $\xi_0(z)=1$.
\item For $k$ varying from $1$ to $n$,
\begin{itemize}
\item For $z$ varying from $0$ to $n-k$,
\begin{itemize}
\item $\xi_{k}(z) \longleftarrow 0$,
\item For $\gamma$ varying from $0$ to $kz$,
$i_{\min} \longleftarrow \max(0,\gamma-z)$, $i_{\max} \longleftarrow \min(\gamma,(k-1)z)$, and
\begin{eqnarray*}
\eta_{k,z}(\gamma) &\longleftarrow & \Indic{i_{\min}\le i_{\max} }\sum_{i=i_{\min}}^{i_{\max}} f(\gamma-i) \bino{z}{\gamma-i} \eta_{k-1, z}(i) \, ,\label{eta7}\\
\xi_{k}(z)&\longleftarrow & \xi_{k}(z) + \eta_{k, z}(\gamma)\sum_{j=0}^{zk-\gamma} \bino{zk-\gamma}{j}(-1)^j\lambda_{j+\gamma} \, .\label{xi3}
\end{eqnarray*}
\end{itemize}
\end{itemize}
\end{itemize}
\end{proposition}

The law of the total defaults' number, with exchangeable direct defaults and exchangeable infections, in the multi-period model, results directly from Theorem~\ref{th3}, and is given by the following algorithm:
\begin{proposition}[Algorithm providing the law of $N_t$ given the coefficients $\xi_k(z)$]\label{algo2}
Once the coefficients $\setindice{\xi_k(z)}{k=0,\dots,n \, ~z=0,\dots,n-k}$ calculated, the law of $N_t$ is given by:
\begin{itemize}
\item Initialization: for $r$ varying from $0$ to $n$, $\Proba{N_0=r}=\Indic{r=0}$.
\item For $t$ varying from $1$ to $T$,
\begin{itemize}
\item For $r$ varying from $0$ to $n$,
\begin{eqnarray*}
\Proba{N_t=r}=\sum_{k=0}^r\Proba{N_{t-1}=k} \bino{n-k}{r-k}\sum_{\gamma=0}^{r-k} \bino{r-k}{\gamma}\sum_{\alpha=0}^{n- k-\gamma} \bino{n-k-\gamma}{\alpha}\mu_{\gamma+\alpha}\sum_{j=0}^{n-r} \bino{n-r}{j}(-1)^{j+\alpha}\xi_{j+r-k-\gamma}(g(k, \gamma)).
\end{eqnarray*}
\end{itemize}
\end{itemize}
\end{proposition}

\subsubsection{Beta-mixture model}
\label{Sec:Chosen_models}

We have chosen a framework enclosing the one by~\citeauthor{DavisLo}, where contagions in a given period are issued from the direct defaults during this period, so that the function $g : \Omega \times \Omega \rightarrow \N$ is given by $g(k,\gamma)=\gamma$. Both direct defaults and infections are  represented by sets of exchangeable random variables. Therefore it becomes necessary to specify the sets $\setindice{\lambda_{j}}{j\in \set{0,\dots,n^2}}$ and $\setindice{\mu_{j}}{j\in \set{0,\dots,n}}$, which characterize these direct defaults and infections.
\modif{We assume that $X_1,\ldots,X_n$ are part of an infinite sequence of exchangeable Bernoulli random variables}. Then, by De Finetti's Theorem (see \cite{Finetti}), there exists a random variable $\Theta_X$, with values in $[0, 1]$ and a cumulative distribution function $F_{\Theta_X}$, such that $\Proba{X^1=1, \dots, X^k=1}=\int_0^1 \theta^k dF_{\Theta_X}(\theta)$. The direct defaults $X^1, \dots, X^n$ behaves as if the Bernoulli parameter $p$ was becoming a hidden common random variable $\Theta_X$. As a consequence, we can write $\mu_j=\Esp{\Theta_X^j}$. We will take $\Esp{\Theta_X}=p$ in order to use the same notations as the ones used by \citeauthor{DavisLo}. We will suppose that the parameter $\sigma_X^2=\Var{\Theta_X}$ is given.
In the same way, we will suppose that variables $\setindice{\Yt^{ij}}{(i,j) \in \Omega^2}$ are Bernoulli random variables, with common hidden parameter $\Theta_Y$. We will write $q=\Esp{\Theta_Y}$ and $\sigma_Y^2=\Var{\Theta_Y}$ and we suppose that these parameters  are  known. We will use for both $\Theta_X$ and $\Theta_Y$ some laws of a same family, distributed on $[0,1]$, where moments are analytically given. For $j \in \N$, coefficients $\mu_j$ and $\lambda_j$ correspond therefore to:
\begin{eqnarray}
\mu_j &=& m(j,p,\sigma_X) \, ,\\
\lambda_j &=& m(j,q, \sigma_Y) \, ,
\end{eqnarray}
where the function $m(j,r,\sigma)$ gives the moment of order $j$ of a random variable $\Theta$ from the considered family of laws, with mean $r$ and standard deviation $\sigma$.\\
We have chosen to use Beta distributions for $\Theta_X$ and $\Theta_Y$, which are parameterized by their mean and standard deviation (it could have been possible to choose other laws, like for example the Kumaraswamy's law).

Moments $m(j,r,\sigma)$ are easy to compute, but we need to consider two separate cases:
\begin{itemize}
\item When $\sigma^2=0$, the hidden random variable $\Theta$ (which may represent $\Theta_X$ or $\Theta_Y$) is a Dirac mass, and we find  the i.i.d. case for which
\begin{eqnarray}
m(j,r,0)=r^j \, .
\end{eqnarray}
\item When $\sigma^2>0$, for a Beta distribution with mean $r=\Esp{\Theta}$ and variance $\sigma^2=\Var{\Theta}$, the parameters $\alpha_{r,\sigma}$ and $\beta_{r,\sigma}$ of the distribution can be easily deduced from $r$ and $\sigma^2$, and we then get all moments of order $j$, $j\in \N$ ($\Gamma$ is here the Euler Gamma function):
\begin{eqnarray}
\alpha_{r,\sigma} &=& r \left( \frac{r(1-r)}{\sigma^2} -1\right) \, ,\\
\beta_{r,\sigma} &=& (1-r) \left( \frac{r(1-r)}{\sigma^2} -1\right) \, , \\
m(j,r,\sigma) &=& \frac{\Gamma(\alpha_{r,\sigma}+j)}{\Gamma(\alpha_{r,\sigma})} \frac{\Gamma(\alpha_{r,\sigma}+\beta_{r,\sigma})}{\Gamma(\alpha_{r,\sigma}+\beta_{r,\sigma}+j)} \, .
\end{eqnarray}
\end{itemize}

\subsection{Effect of model parameters on the dynamics of loss distributions}
\label{Sec:Effect_of_parameters}


All numerical values that we used for the parameters in this section remain to be specified. We grouped them into the following Definition~\ref{defmodele}.
\begin{definition}\label{defmodele}
We define $4$ reference models. These models have some shared characteristics: they all consider $10$ firms ($n=10$), on a time interval divided into 10 periods  ($T=10$), with a direct default probability $p=0.1$ and an infection probability $q=0.2$.
The four models are distinguished on the nature of direct defaults or infections, i.i.d. or not (depending on the two parameters $\sigma_X$ and $\sigma_Y$) and on the infections number that is required to cause a default, depending on the function $f$.
The specificities of the four models are presented hereafter:
\begin{itemize}
\item \textit{model 1}: $\sigma_X=0$, $\sigma_Y=0$, $f(x)=\Indic{x\ge 1}$ (i.i.d. case, one required contamination).
\item \textit{model 2}: $\sigma_X=0$, $\sigma_Y=0$, $f(x)=\Indic{x\ge 2}$ (i.i.d. case, two required contaminations).
\item \textit{model 3}: $\sigma_X=0.2$, $\sigma_Y=0.2$, $f(x)=\Indic{x\ge 1}$ (exchangeable case, one required contamination).
\item \textit{model 4}: $\sigma_X=0.2$, $\sigma_Y=0.2$, $f(x)=\Indic{x\ge 2}$ (exchangeable case, two required contaminations).
\end{itemize}
\end{definition}

\modif{Starting from the same model assumptions, we have checked that, the coefficients $\xi_k(z)$, $k \in \set{0,\dots,n}$, $z \in \set{0,\dots,n-k}$ computed either from formula 1 or formula 2 of Proposition~\ref{prop1} give the same loss distribution $\Proba{N_t=k}, k\in \Omega$ when $f(x)=\Indic{x\ge 1}$.} In this last case, if  furthermore $\sigma_X=\sigma_Y=0$ and $t=1$, we have checked that we get by these two formulas the same values for $\Proba{N_t=k}, k\in \Omega$ as the one computed with Davis and Lo's result.\\

\begin{figure}[h!]
\includegraphics[ angle=-90, width=0.5\linewidth]{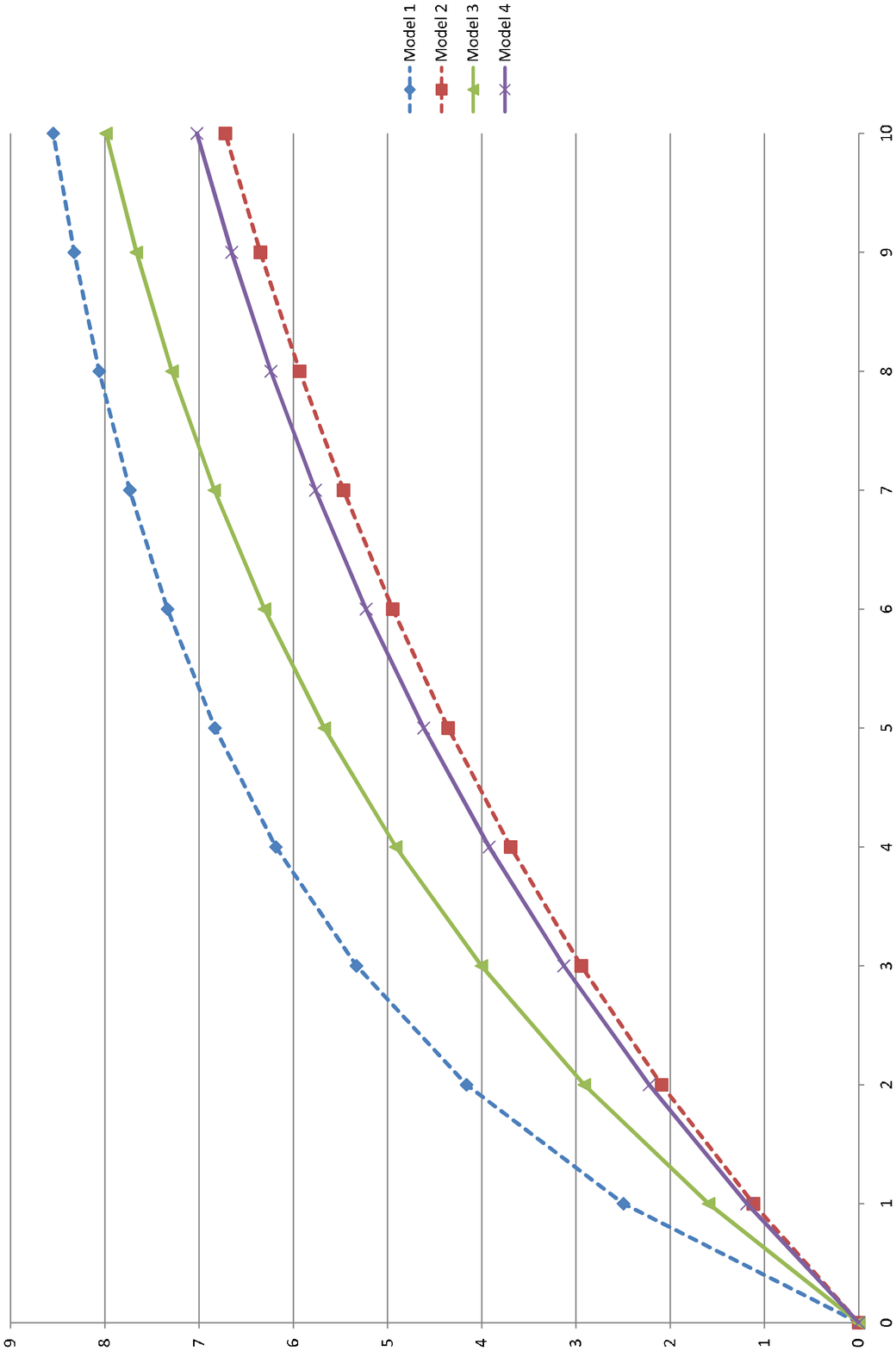}
\includegraphics[angle=-90, width=0.5\linewidth]{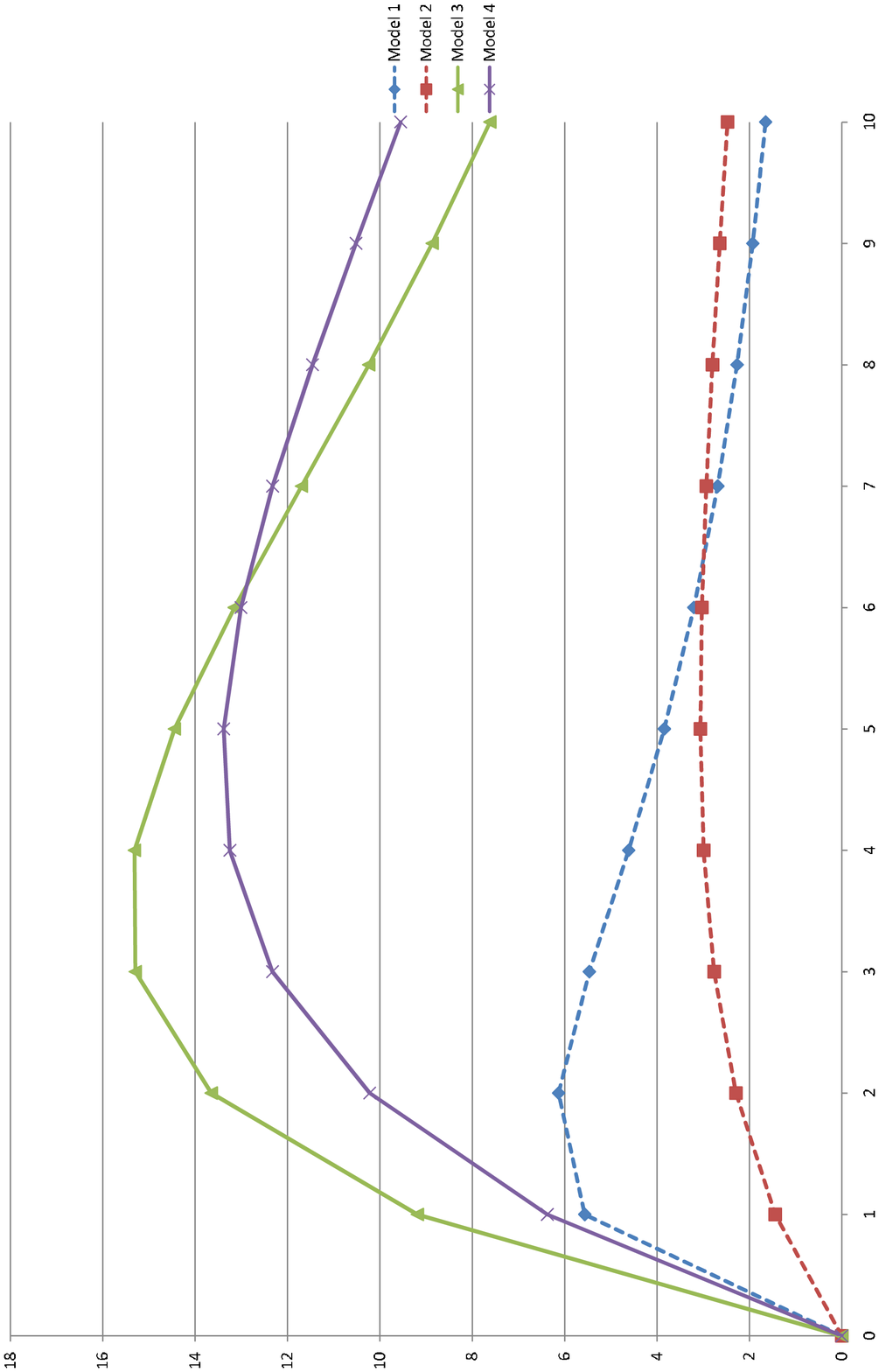}
\caption[]{\small{Evolution of $\Esp{N_t}$ (left) and $\Var{N_t}$ (right) as a function of $t$, for all the models described in Definition~\ref{defmodele}.
The curves  plotted with dotted lines correspond to the case where direct defaults are i.i.d.}}
\label{GrapheENt_VNt}
\end{figure}

The evolutions of expectation and variance of $N_t$ as a function of $t$, for all of the four models that are described in Definition~\ref{modele}, are illustrated in Figures~\ref{GrapheENt_VNt}. \modif{We observe that the mean of the number of defaults distribution is an increasing function of time. This is obviously consistent with the situation where more defaults are expected in a larger period of time.} In the exchangeable case (models 3 and 4), direct defaults as well as infections arise from probabilities that depend on hidden factors. If the distribution of $N_t$ is intuitively more dispersed, the impact of such a dispersion on the mathematical expectation is not trivial. For the mean of $N_t$, we can observe on Figure~\ref{GrapheENt_VNt} (left side) that the expectation of $N_t$ is \modif{only slightly} modified when choosing exchangeable random variables, but this change is very small in the case where two contagions are required to cause a default, i.e., when $f(x)=\Indic{x \ge 2}$ (models 2 and 4). \modif{This suggests that when the contagion effect is weakened as this is the case here, the mean of the loss distribution is mainly explained by the mean of the direct defaults, which are the same in models 2 and 4.} When considering the variance of $N_t$ plotted in Figure~\ref{GrapheENt_VNt} (right side), we observe more evident results. \modif{The variance of the loss distribution is a hump-shaped function of time. It is quite intuitive that the dispersion level increases as time goes by since the loss process can reach a larger part of its support with higher probabilities. However, when the expected number of defaults attains a critical threshold, the number of expected surviving names then decreases, leading at some point in time to a decrease in the dispersion level of the loss distribution. Let us remark that this occurs earlier for models 1 and 3 for which the expected number of defaults is greater.} Moreover, the presence of a variation in the hidden random variables $\Theta_X$ and $\Theta_Y$ naturally implies a larger increase of the variance of $N_t$.\\

\begin{figure}[h!]
\includegraphics[angle=-90, width=0.5\linewidth]{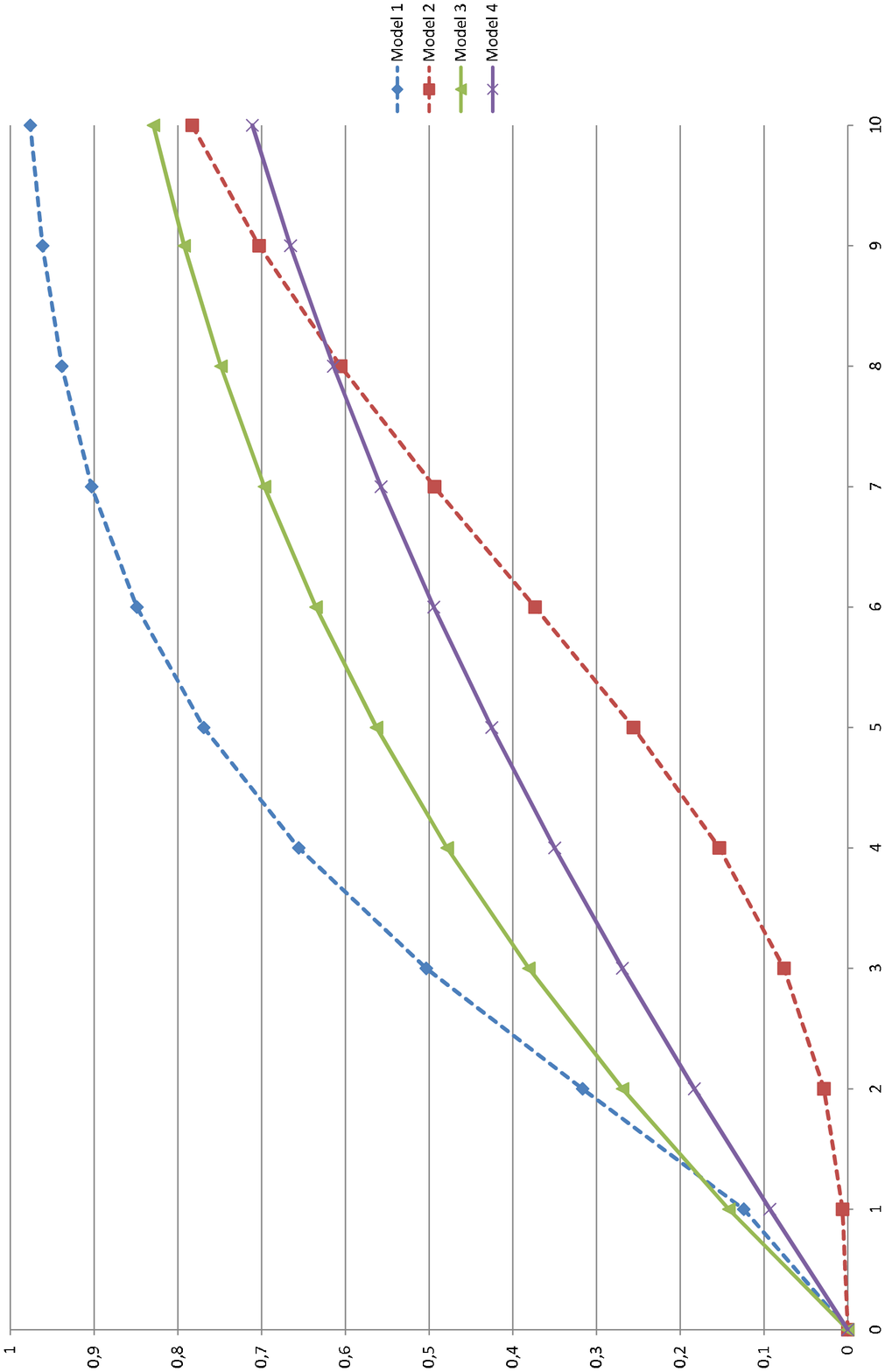}
\includegraphics[angle=-90,  width=0.5\linewidth]{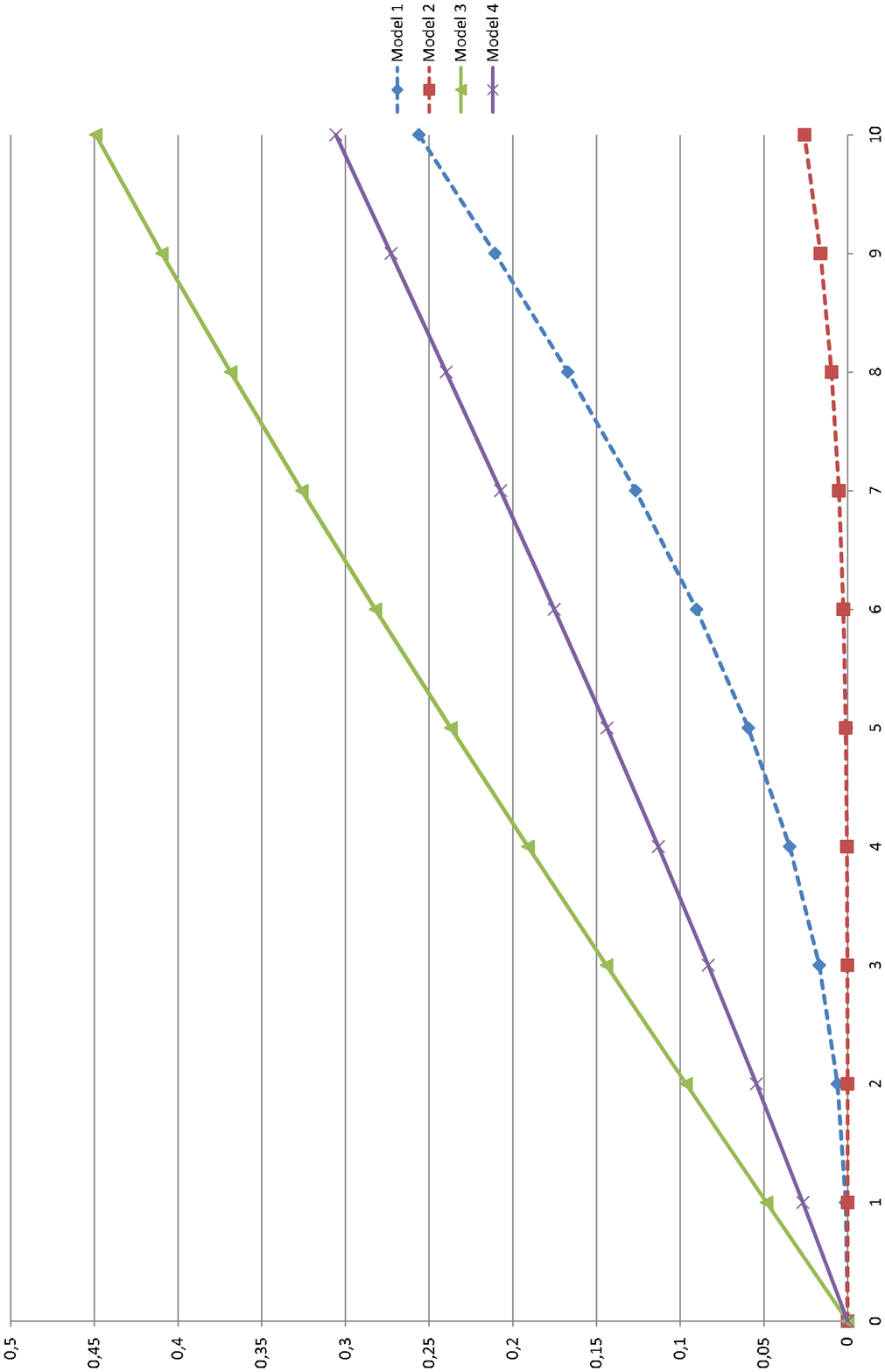}
\caption[]{ \small{Evolution of $\Proba{N_t\ge 6}$ and $\Proba{N_t=10}$ as a function of $t$, for all the models described in Definition~\ref{defmodele}. The curves  plotted with dotted lines correspond to the case where direct defaults are i.i.d.}}
\label{GrapheSN5t_SN9t}
\end{figure}

Figure~\ref{GrapheSN5t_SN9t} shows the evolution of $\Proba{N_t \ge k}$, for respective default number $k=6$ (left side) and $k=10$ (right side), among the ten considered firms ($n=10$). With the chosen parameters, we observe that the impact of exchangeability is not systematically the same. The growth of the volatility of $N_t$ does not always lead to the increase of $\Proba{N_t \ge k}$, especially when this last probability is larger in the independence case. When we consider less frequent events, as illustrated on the right side of Figure~\ref{GrapheSN5t_SN9t} where the probability that all firms default is plotted, we find, on the contrary, that the impact of exchangeability is far more explicit. This seems logical, since the increase of the dispersion level of $N_t$, due to hidden parameters $\Theta_X$ and $\Theta_Y$, \modif{tends to emphasize the tail of the distribution}. \modif{This is clearly a required model behavior especially when one has in mind some applications to the pricing of synthetic CDO tranches. Indeed, one can think of the base correlation breakdown in March 2008, when no implied correlation can be found for CDX super-senior tranches.}

\subsection{Calibration on liquid CDO tranche quotes}
\modif{
This section examines the fit of the model to tranche spreads of the 5-years iTraxx Europe main index at two fixed points in time, namely August 31, 2005 and March 31, 2008. These two dates have been chosen to facilitate comparison of model calibration before and during the credit crisis. Let us recall that synthetic CDOs are structured products based on an underlying portfolio of reference entities subject to credit risk. It allows investors to sell protection on specific risky portions or tranches of the underlying credit portfolio depending on their desired risk-profile. We concentrate our numerical applications on the most liquid segment of the market, namely CDO tranches written on standard CDS indexes such as the iTraxx Europe main index. As illustrated in Figure \ref{Schema:Structure_CDO_Standardise_iTraxx}, the iTraxx Europe main index contains 125 investment grade CDS, written on large European corporations. Market-makers of this index have also agreed to quote standard tranches on these portfolios from equity or first loss tranches to the most senior tranches. Each tranche is defined by its attachment point which is the level of subordination and its detachment point which is the maximum loss of the underlying portfolio that would result in a full loss of tranche notional. The first-loss 0-3\% equity tranche is exposed to the first several defaults in the underlying portfolio. This tranche is the riskiest as there is no benefit of subordination but it also offers high returns if no default occurs. The junior mezzanine 3-6\% and the senior mezzanine
6-9\% tranches are less immediately exposed to the portfolio defaults but the
premium received by the protection seller is smaller than for the equity tranche.
The 9-12\% tranche is the senior tranche, while the 12-20\% tranche is the low-risk super senior piece.

\begin{figure}[ht]
\centering
        \centering
        \includegraphics[bb=116   512   456   754, scale=0.6]{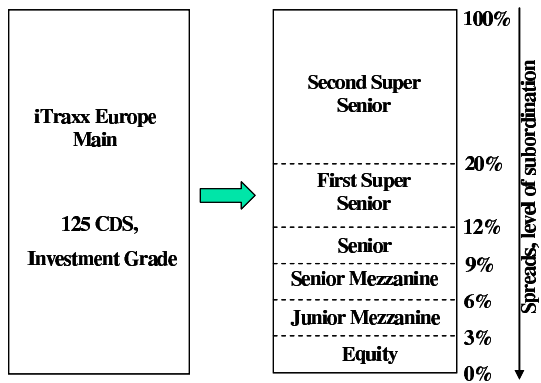}
        \caption{Standardized CDO tranches on iTraxx Europe Main.
            \label{Schema:Structure_CDO_Standardise_iTraxx}
        }
\end{figure}

Since individual credit exposures are all the same among names in a standard CDS index, the key quantity involved in the pricing of standard tranches is the cumulative loss per unit of nominal exposure defined by
$$
L_t = \sum_{i=1}^n (1-R_i) Z_{t}^i,
$$
where $R_i$ denotes the recovery rate associated with name $i$. We recall that $Z_{t}^i$ referred to as the time-$t$ default indicator of name $i$, the latter quantity being defined by Equations (\ref{Eq:default_indicators_multi_period_1}) and  (\ref{Eq:default_indicators_multi_period_2}) as in the previously presented multi-period extension of the Davis and Lo's model. The cash-flows associated with a synthetic CDO tranche with attachment point $a$ and detachment point $b$ ($a$ and $b$ in percentage) are driven by losses that affect the tranche, i.e.,
 $$ L_t^{(a,b)}  = \left( {L_t  - a} \right)^ +   - \left( {L_t  - b} \right)^ +. $$
We refer the reader to \cite{Cousin_Laurent2008} for a detailed description of cash-flows associated with synthetic CDO tranches. The model that we consider for the calibration of CDO tranche quotes belongs to the class of models presented in section \ref{Sec:Chosen_models}. More specifically, we assume that, for any $t\geq 0$, direct defaults $X_t^i$, $i=1,\ldots,n$ are Bernoulli mixtures where the common hidden parameter is a Beta-distributed random variable with mean $p$ and variance $\sigma_X^2$. For the sake of parsimony and in order to avoid numerical difficulties when $n$ is large, we consider a version of the model where only three parameters are needed. We consider that, for any $t\geq 0$, contagion events $Y_t^{i,j}$, $1 \leq i,j\leq n$ are independent Bernoulli random variables with the same mean $q$ ($\sigma_Y=0$). We consider here the case where only one contagion is required to cause a default. Using the latter restrictions, the discrete-time contagion model is stationary and entirely described by the vector of annual scaled parameters $\alpha = (p, \sigma_X, q)$. Let us recall that the computation of CDO tranche spreads only involves the expectation of tranche losses at several time horizons. In the case where recovery rates are the same across names and equal to a constant $R$, it is straightforward to remark that the current cumulative loss is merely proportional to the current number of defaults. The algorithms described in Subsection \ref{Seq:Algo} can then be used properly to compute CDO tranche spreads.\\

Let us denote by $\tilde{s}_0$, $\tilde{s}_1$, $\tilde{s}_2$, $\tilde{s}_4$, $\tilde{s}_5$, $\tilde{s}_6$ the market spreads associated with (respectively) the CDS index, [0\%-3\%], [3\%-6\%], [6\%-9\%], [9\%-12\%] and [12\%-20\%] standard iTraxx tranches and by $s_0(\alpha)$, $s_1(\alpha)$, $s_2(\alpha)$, $s_4(\alpha)$, $s_5(\alpha)$, $s_6(\alpha)$, the corresponding spreads generated by the contagion model using the vector of parameters $\alpha$. The calibration process aims at finding out the optimal parameter set $\alpha^*=(p^*, \sigma_X^*, q^*)$ which minimizes the following least-square objective function
$$
RMSE(\alpha) = \sqrt{\frac{1}{6}\sum_{i=1}^6\left(\frac{\tilde{s}_i- s_i(\alpha)}{\tilde{s}_i}\right)^2}.
$$
For both data sets, in order to analyze the calibration efficiency in a deeper way, we have compared the global calibration with three alternative ones, where some of the available market spreads were excluded from the fitting. Here are the calibration procedures we have considered:

\begin{itemize}
\item Calibration 1: All available market spreads are included in the fitting,
\item Calibration 2: The equity [0\%-3\%] tranche spread is excluded,
\item Calibration 3: Both equity [0\%-3\%] tranche and CDS index spreads are excluded,
\item Calibration 4: All tranche spreads are excluded except equity tranche and CDS index spreads.
\end{itemize}

In all calibrations the interest rate is set to 3\%, the payment frequency is quarterly and the recovery rate is $R=40\%$.  We provide in Table \ref{tab:calib_results_2005} and Table \ref{tab:calib_param_2005} model spreads and optimal parameters resulting from the four benchmark calibration processes performed on August 31$^{\text{st}}$, 2005 iTraxx quotes. Table \ref{tab:calib_results_2008} and Table \ref{tab:calib_param_2008} present the same types of outcomes but for calibrations on March 31$^{\text{st}}$, 2008 iTraxx quotes.\\

As can be seen from Tables \ref{tab:calib_results_2005} and \ref{tab:calib_results_2008},
the calibration of the three parameters on all market spreads is rather disappointing. This is not surprising, especially given the poor calibration performance of standard factor models during the crisis, when the fit is achieved on all tranches and index quotes. Moreover, we observed that calibration results are fairly dependent on the initial starting point specified in the minimization routine. We do not guarantee that the presented outcomes correspond to a global minimum of the objective function. However, one can note that the calibration error decreases when we subsequently exclude the equity tranche quote and the index quote in Calibrations 2 and 3. Interestingly enough, the fit on 2008 data is nearly perfect in Calibration 3 (see Table \ref{tab:calib_results_2008}) compared with Calibration 3 on 2005 data (see Table \ref{tab:calib_results_2005}). Unsurprisingly, as illustrated by results from Calibration 4, the fit on equity and index spreads only is perfect. We have checked that this is actually the case for all tranches when they are jointly fitted with the index. This can be seen as a fundamental required behavior of the model since we try to fit three parameters on two market quotes. Let us recall however that the base correlation framework had some difficulties to fit the super-senior tranches in the same period (March 2008). As for the optimal parameters, we observe that the one-year mean and dispersion level of direct default events (parameters $p^*$ and $\sigma_X^*$) are higher in 2008 calibrations compared with 2005 calibrations. This can be explained by the drastic shift in credit spreads between these two periods. However, we can note that the optimal contagion parameter $q^*$ has the same order of magnitude in both periods except for Calibration 1 in 2008 where it appeared to be insignificant.\\

\begin{table}[htbp]
  \centering
    \begin{tabular}{rccccccc}
    \hline
    \hline
          & 0\%-3\% & 3\%-6\% & 6\%-9\% & 9\%-12\% & 12\%-20\% & index & RMSE \\
    \hline
    Market quotes & 24    & 81    & 27    & 15    & 9     & 36    & - \\
    Calibration 1 & 20    & 114   & 7     & 1     & 1     & 29    & 0.64 \\
    Calibration 2 & -     & 62    & 32    & 18    & 6     & 8     & 0.41 \\
    Calibration 3 & -     & 55    & 29    & 18    & 7     & -     & 0.22 \\
    Calibration 4 & 24    & -     & -     & -     & -     & 36    & 0 \\
    \hline
    \hline
    \end{tabular}
    \caption{iTraxx Europe main, August 31$^{\text{st}}$, 2005. The market and model spreads (in bp) in the four calibrations and the corresponding root mean square errors. The [0\%-3\%] spread is quoted in \%. All maturities are for five years.}
  \label{tab:calib_results_2005}
\end{table}

\begin{table}[htbp]
  \centering
    \begin{tabular}{rccc}
    \hline
    \hline
          & $p^*$     &  $\sigma_X^*$ & $q^*$ \\
    \hline
    Calibration 1 & 0.0016 & 0.0015 & 0.0626 \\
    Calibration 2 & 0.0007 & 0.0133 & 0.0400 \\
    Calibration 3 & 0.0001 & 0.0025 & 0.3044 \\
    Calibration 4 & 0.0014 & 0.002 & 0.1090 \\
    \hline
    \hline
    \end{tabular}
    \caption{iTraxx Europe main, August 31$^{\text{st}}$, 2005. Optimal parameters $\alpha^*=(p^*,\sigma_X^*,q^*)$ resulting from the four calibration procedures ($\sigma_Y=0$).}
  \label{tab:calib_param_2005}
\end{table}

\begin{table}[htbp]
  \centering
    \begin{tabular}{rccccccc}
    \hline
    \hline
          & 0\%-3\% & 3\%-6\% & 6\%-9\% & 9\%-12\% & 12\%-20\% & index & RMSE \\
    \hline
    Market quotes & 40    & 480   & 309   & 215   & 109   & 123   & - \\
    Calibration 1 & 28    & 607   & 361   & 228   & 95    & 75    & 0.25 \\
    Calibration 2 & -     & 505   & 330   & 228   & 112   & 68    & 0.20 \\
    Calibration 3 & -     & 478   & 309   & 215   & 109   & -     & 0.002 \\
    Calibration 4 & 40    & -     & -     & -     & -     & 123   & 0 \\
    \hline
    \hline
    \end{tabular}
    \caption{iTraxx Europe main, March 31$^{\text{st}}$, 2008. The market and model spreads (in bp) in the four calibrations and the corresponding root mean square errors. The [0\%-3\%] spread is quoted in \%. All maturities are for five years.}
  \label{tab:calib_results_2008}
\end{table}

\begin{table}[htbp]
  \centering
    \begin{tabular}{rccc}
    \hline
    \hline
          & $p^*$     & $\sigma_X^*$ & $q^*$ \\
    \hline
    Calibration 1 & 0.0124 & 0.0886 & 0 \\
    Calibration 2 & 0.0056 & 0.0518 & 0.0400 \\
    Calibration 3 & 0.0012 & 0.012 & 0.2688 \\
    Calibration 4 & 0.0081 & 0.0516 & 0.0589 \\
    \hline
    \hline
    \end{tabular}
    \caption{iTraxx Europe main, March 31$^{\text{st}}$, 2008. Optimal parameters $\alpha^*=(p^*,\sigma_X^*,q^*)$ resulting from the four calibration procedures ($\sigma_Y=0$).}
  \label{tab:calib_param_2008}
\end{table}

To summarize, in both data sets, the best
calibrations were obtained when both the CDS index and equity tranche spreads were excluded. Moreover, all calibrations tested on the 2008 data  seem to outperform calibrations achieved on the 2005 data. Even though additional investigations have to be achieved, the latter result suggests that the dynamic feature together with the contagion effect seem to have a significant impact  on pricing performance, especially during the recent distressed period.
}

\section{Conclusion}

\modif{
In this paper, we examine a multi-period extension of the original Davis and Lo's contagion model. The model allows for both spontaneous or direct defaults and defaults resulting from a chain reaction where defaulted names at the beginning of a given period can infect other firms by contagion at the end of this period. Moreover, the number of defaulted names required to trigger a contagion effect can be specified as a model input. At last, random variables governing direct defaults and contaminations may depend on common macroeconomic factors. In a rather general setting, we provide a recursive algorithm for the computation of the number of defaults distribution at several time horizons.}\\

\modif{
 In numerical applications, we restrain ourselves to the case where, between each time interval, both direct defaults and contamination events are described by mixtures of Bernoulli variables mixed with a Beta-distributed random factor. We first illustrate the effect of model parameters on the dynamics of loss distributions. In particular, we investigate the factors dispersion impact and the contagion effect on the variance and the quantiles of the loss distribution. We then consider the fit of the model parameters on iTraxx data before and during the credit crisis. This allows to exploit the dynamic feature of the model and illustrate its tractability when the number of reference entities is large (and equal to 125). In both data sets, the best calibrations were obtained when both the CDS index and equity tranche spreads were excluded with a perfect fit in March 2008. In that case, we can remark that the contagion probability is meaningful. Surprisingly, all calibrations tested on the 2008 data seem to outperform calibrations achieved on the 2005 data. Even though additional investigations have to be achieved, the latter results suggest that the dynamic feature together with the contagion effect seem to have a significant impact on pricing performance, especially during the recent distressed period.}\\

\modif{
Eventually, this study paves the way to additional applications such as the management of counterparty risk arising in CDS or CDO tranches transactions.
}\\
\FloatBarrier
\section{Appendix}
In this part, we give the proofs of Proposition~\ref{prop1} and Lemma~\ref{lemme2}

\textbf{Proof of Proposition \ref{prop1}}
\begin{enumerate}
\item

In the particular case of infinite exchangeability assumption, the first part of Proposition \ref{prop1} is a consequence of De Finetti's Theorem (see \cite{Finetti}). Indeed, since the variables $\set{\Yt^{ij}, (i, j)\in\Omega^2}$ are exchangeables, there exists a random variable $\Theta_Y\in [0, ~1]$, such that, conditionally to $\Theta_Y$, random variables $\set{\Yt^{ij}, (i, j)\in\Omega^2}$ are i.i.d., Bernoulli distributed with parameter $\Theta_Y$. As a consequence
\begin{align*}
\xi_{k,t}(z)=&\Esp{\Proba{\displaystyle\sum_{\substack{j\in F_t \\ card(F_t)=z}} \Yt^{ji}\ge 1 | \Theta_Y}^k}=\Esp{(1-(1-\Theta_Y)^{z})^k}=\sum_{i=0}^{k} \bino{k}{i}(-1)^i\Esp{(1-\Theta_Y)^{iz}}.
\end{align*}
Noting that $\Esp{(1-\Theta_Y)^{iz}}=\Esp{\displaystyle\sum_{\substack{j\in A \\ card(A)=zi}} \Yt^{j1}=0}$, we conclude using Lemma \ref{Waring}.

\modif{In the  case of finite exchangeability assumption, De Finetti's Theorem may not be applied. That's why it is necessary to use another reasoning :}

Let $m\geq 1$. Let us define for $k \in \set{1, \dots, m-1}$:
\begin{align*}
\xi_{m,t}\ordre k(z) &= \Proba{\sum_{j\in F_t} \Yt^{j1}\geq 1 \cap ... \sum_{j\in F_t} \Yt^{j, m-k}\geq 1\cap \sum_{j\in F_t} \Yt^{j, m-k+1 }=0 \cap \dots \sum_{j\in F_t} \Yt^{j, m} = 0\sachant card(F_t)=z} \, ,\\
\xi_{m,t}\ordre m(z) &= \Proba{\sum_{j\in F_t} \Yt^{j1}=... =\sum_{j\in F_t} \Yt^{j, m}=0\sachant card(F_t)=z}.
\end{align*}
Since the variables $\set{\Yt^{ij}, (i, j)\in\Omega^2}$ are exchangeables, we see that for $k \in \set{1, \dots, m}$,
$$\xi_{m,t} \ordre k + \xi_{m,t} \ordre{k-1} = \xi_{m-1,t} \ordre{k-1} \, .$$ It follows that for  all $k$ ($1\leq k \leq j$) :
$$\xi_{j,t} \ordre{k-1} = \xi_{k-1,t} \ordre{k-1} - \sum_{m=k}^j \xi_{m,t} \ordre k \, .$$
By induction on $k=1, \dots, m$, we get:
\begin{align*}
\xi_{k,t}(z)=&1- \bino k1 \xi_{1,t} \ordre 1 (z) + \bino k2 \xi_{2,t} \ordre 2 (z) + ...+(-1)^k \bino kk \xi_{k,t} \ordre k (z).
\end{align*}
Since the variables $\set{\Yt^{ij}, (i, j)\in\Omega^2}$ are exchangeables, then for all $r$ ($1\leq r\leq k$) , we conclude using Lemma \ref{Waring} :
$$\xi_{r,t} \ordre r=\Proba{\sum_{j=1}^{rz} \Yt^{\sigma(j)}=0} = \sum_{\alpha=0}^{zj} \bino{zj}{\alpha} (-1)^{\alpha} \lambdat{\alpha} \, .$$

\item Let $f_z^{-1}=\set{i\in\Omega : i\leq z, ~f(i)=1}$. Then
$$\xi_{k,t}(z)=\displaystyle\sum_{\gamma_1,\dots,\gamma_k\in f_z^{-1}}\Proba{\sum_{j\in F_t} \Yt^{j1}=\gamma_1 \cap ... \cap \sum_{j\in F_t} \Yt^{jk}=\gamma_k\sachant card(F_t)=z}.$$
Furthermore random variables $\set{\Yt^{ji}, ~(j, i)\in\Omega^2}$ are exchangeable. So, using Remark~\ref{rem2}
 \begin{align*}
 \xi_{k,t}(z)&=\displaystyle\sum_{\gamma_1,\dots,\gamma_k\in f_z^{-1}}\frac{ \bino{z}{\gamma_1}\dots \bino{z}{\gamma_k}}{\bino{kz}{\gamma_1+\dots+\gamma_k}}\Proba{\sum_{j=1}^{kz} \Yt^{\sigma(j)}=\gamma_1+\dots +\gamma_k}\\
 &=\displaystyle\sum_{\gamma=0}^{kz}\sum_{\substack{\gamma_1+\dots +\gamma_k=\gamma \\ \gamma_1,\dots, \gamma_k\in\set{0,\dots ,z}}}f(\gamma_1)\dots f(\gamma_k)\frac{ \bino{z}{\gamma_1}\dots \bino{z}{\gamma_k}}{\bino{kz}{\gamma}}\Proba{\sum_{j=1}^{kz} \Yt^{\sigma(j)}=\gamma}.
 \end{align*}
From Lemma~\ref{Waring}
 $$\Proba{\sum_{j=1}^{kz} \Yt^{\sigma(j)}=\gamma}= \bino{kz}{\gamma}\sum_{j=0}^{kz-\gamma} \bino{kz-\gamma}{j}(-1)^j\lambda_{j+k, t}.$$
 As a consequence
 $$\xi_{k,t}(z)=\sum_{\gamma=0}^{zk}\eta_{k, z}(\gamma)\sum_{j=0}^{zk-\gamma} \bino{zk-\gamma}{j}(-1)^j\lambda_{j+\gamma, t},$$
where $\eta_{k, z}(\gamma)=\displaystyle\sum_{\substack{\gamma_1+\dots +\gamma_k=\gamma \\ \gamma_1,\dots, \gamma_k\in\set{0,\dots ,z}}}f(\gamma_1)\dots f(\gamma_k) \bino{z}{\gamma_1}\dots \bino{z}{\gamma_k}.$
\\
A simple calculation leads us to the following reccurence relationship :
$$\eta_{k, z}(\gamma)=\displaystyle\sum_{\substack{\gamma_1\in \set{0, \dots, z}\\\gamma_1\leq\gamma}}f(\gamma_1) \bino{z}{\gamma_1} \eta_{k-1, z}(\gamma-\gamma_1).$$
\EndProof
\end{enumerate}
\textbf{Proof of Lemma~\ref{lemme2}}
From the definition of conditional probability $$\Proba{N_t=r\sachant N_{t-1}=k}=\frac{\Proba{N_t=r, N_{t-1}=k}}{\Proba{N_{t-1}=k}}.$$

Under Assumptions~\ref{h2} and~\ref{h4}, the random variables $\set{Z^i_t, ~i\in\Omega, t\in\set{1, \dots, T}}$ are exchangeable. On the one hand, using two times Remark~\ref{rem2}, we establish that $\Proba{N_t=r, N_{t-1}=k}$ is equal to
\begin{align*}
&\bino nk \Proba{N_t=r, Z^1_{t-1}=\dots=Z^k_{t-1}=1, Z^{k+1}_{t-1}=\dots=Z^n_{t-1}=0}\\
=& \bino nk \Proba{N_t=r, Z^1_{t}=\dots=Z^k_{t}=Z^1_{t-1}=\dots=Z^k_{t-1}=1, Z^{k+1}_{t-1}=\dots=Z^n_{t-1}=0}\\
=& \bino nk \bino{n-k}{r-k}\Proba{Z^1_{t}=\dots=Z^r_{t}=Z^1_{t-1}=\dots=Z^k_{t-1}=1, Z^{r+1}_{t}=\dots=Z^n_{t}=Z^{k+1}_{t-1}=\dots=Z^n_{t-1}=0}\\
=&\bino nk \bino{n-k}{r-k}\Proba{\Theta_t=\theta_t, \Theta_{t-1}=\theta_{t-1}}.
\end{align*}
On the other hand, using the same remark, $\Proba{N_{t-1}=k}=\bino nk \Proba{ \Theta_{t-1}=\theta_{t-1}}$, which concludes the proof.
\EndProof



\begin{thebibliography}{99}

\signesmarge{$\star$}










\bibitem[Arnsdorf and Halperin(2007)]{Arnsdorf.Halperin2007} \rien Arnsdorf, M., Halperin, I. (2007) \it BSLP: Markovian bivariate spread-loss model for portfolio credit derivatives, \rm  working paper, JP Morgan.

\bibitem[Azizpour and Giesecke(2008)]{Azizpour.Giesecke2008} \rien Azizpour, S., Giesecke, K. (2008) \it Self-exciting corporate defaults: contagion vs. frailty, \rm  working paper, Stanford University.

\bibitem[Boissay(2006)]{Boissay2006} \rien Boissay, F. (2006) \it Credit chains and the propagation of financial distress, \rm  working paper series 573, European Central Bank.

\bibitem[Cont, Deguest and Kan(2009)]{Cont.Deguest.ea2009} \rien Cont, R., Deguest, R., Kan, Y.H. (2009) \it Default intensities implied by CDO spreads: inversion formula and model calibration, \rm  working paper, Columbia University.

\bibitem[Cont and Minca(2008)]{Cont.Minca2008} \rien Cont, R., Minca, A. (2008) \it Recovering portfolio default intensities implied by {CDO} quotes, \rm  working paper, Columbia University.

\bibitem[Cousin and Laurent(2008)]{Cousin_Laurent2008} \rien Cousin, A., Laurent, J.-P. (2008) \it An overview of factor models for pricing CDO tranches, \rm  Frontiers In Quantitative Finance, Ed. R. Cont, Wiley Finance.

\bibitem[Cousin, Jeanblanc and Laurent(2009)]{Cousin_Jeanblanc_Laurent2009} \rien Cousin, A., Jeanblanc, M., Laurent, J.-P. (2009) \it Hedging CDO tranches in a Markovian environment, \rm  book chapter, working paper.

\bibitem[Das, Duffie, Kapadia and Saita(2007)]{Das.Duffie.ea2007} \rien Das, S.R., Duffie, D., Kapadia, N., Saita, L. (2007) \it Common failings: how corporate defaults are correlated, \rm  Journal of Finance 62(1), 93-117.

\bibitem[Davis and Lo(2001)]{DavisLo} \rien Davis, M., Lo, V. (2001) \it
Infectious Defaults, \rm  Quantitative Finance 1, 382-387.

\bibitem[De Finetti(1931)]{Finetti} \rien De Finetti, B. (1931) \it Funzione caratteristica di un fenomeno
aleatorio, \rm Atti della R. Academia Nazionale dei Lincei, Serie 6.
Memorie, Classe di Scienze Fisiche, Mathematice e Naturale, 251-299.

\bibitem[Denuit, Dhaene and Ribas(2001)]{Denuit1} \rien Denuit, M., Dhaene, J., Ribas, C. (2001) \it Does positive dependence between individual risks increase stop-loss premiums?, \rm Insurance: Mathematics and Economics 28, 305-308.

\bibitem[Denuit, Lefèvre and Utev(2002)]{Denuit} \rien Denuit, M., Lefèvre, Cl., Utev, S. (2002) \it Measuring the impact of dependence between claims occurences, \rm Insurance: Mathematics and Economics 30(1), 1-19.

\bibitem[De Koch, Kraft and Steffensen(2007)]{DeKoch.Kraft2007} \rien De Koch, J., Kraft, H., Steffensen, M. (2007) \it CDOs in chains, \rm working paper, University of Kaiserslautern.

\bibitem[Egloff, Leippold and  Vanini(2007)]{Egloff} \rien Egloff, D.,  Leippold, M., Vanini, P. (2007) \it A simple model of credit contagion, \rm Journal of Banking \& Finance, Elsevier, vol. 31(8),  2475-2492.


\bibitem[Epple, Morgan and Schloegl(2007)]{Epple.Morgan.ea2007} \rien Epple, F., Morgan, S., Schloegl, L. (2007) \it Joint distributions of portfolio losses and exotic portfolio products, \rm International Journal of Theoretical and Applied Finance 10(4), 733-748.

\bibitem[Feller(1968)]{Feller} \rien Feller, W. (1968) \it An Introduction to Probability Theory and Its Applications, \rm Volume I, John-Wiley \& Sons.

\bibitem[Frey and Backhaus(2007)]{Frey} \rien Frey, R., Backhaus, J. (2007) \it A Dynamic hedging of syntetic CDO tranches with spread and contagion risk, \rm working paper, University of Leipzig.

\bibitem[Frey and Backhaus(2008)]{Frey2008} \rien Frey, R., Backhaus, J. (2008) \it Pricing and hedging of portfolio credit derivatives with interacting
	default intensities, \rm International Journal of Theoretical and Applied Finance 11(6), 611-634.

\bibitem[Frey and Runggaldier(2008)]{FreyRunggaldier2008} \rien Frey, R., Runggaldier, W. (2008)\it Pricing credit derivatives under incomplete information: a nonlinear filtering approach, \rm{Finance and Stochastics}, Forthcoming.

\bibitem[Gerber(1995)]{Gerber} \rien Gerber, H.U. (1995) \it Life Insurance Mathematics (2nd ed.), \rm Springer-Verlag.

\bibitem[Giesecke and Weber(2004)]{Giesecke} \rien  Giesecke, K.,  Weber, S. (2004) \it Cyclical correlations, credit contagion, and portfolio losses, \rm Journal of Banking \&  Finance 28, 3009-3036.

\bibitem[Graziano and Rogers(2006)]{GrazianoRogers2006} \rien Graziano, G., Rogers, C. (2006) \it A dynamic approach to the modelling of correlation credit derivatives using Markov chains, \rm working paper, Statistical Laboratory, University of Cambridge.

\bibitem[Herbertsson(2007)]{Herbertsson2007} \rien Herbertsson, A. (2007) \it Pricing synthetic CDO tranches in a model with default contagion
	using the matrix-analytic approach, \rm Journal of Credit Risk 4(4), 3-35.

\bibitem[Herbertsson and Rootz{\'e}n(2006)]{Herbertsson.Rootz'en2006} \rien Herbertsson, A. and Rootz{\'e}n, H. (2006) \it Pricing k-th to default swaps under default contagion, the matrix-analytic
	approach, \rm Journal of Computational Finance 12(1), 49-78.

\bibitem[Jarrow and Yu(2001)]{JarrowYu} \rien Jarrow, R., Yu, F. (2001) \it Counterparty risk and the pricing of defaultable securities, \rm Journal of Finance 56, 1765-1799.

\bibitem[Jorion and Zhang(2007)]{JorionZhang2007} \rien Jorion, P., Zhang, G. (2007) \it Good and bad credit contagion: Evidence from credit default swaps, \rm Journal of Financial Economics 84(3), 860-883.

\bibitem[Jorion and Zhang(2009)]{JorionZhang2009} \rien Jorion, P., Zhang, G. (2009) \it Information transfer effects of bond rating downgrades, \rm{The Financial Review}, Forthcoming.

\bibitem[Kraft and Steffensen(2007)]{Kraft} \rien Kraft, H.,  Steffensen, M. (2007) \it Bankruptcy, counterparty risk, and contagion, \rm Review of Finance 11,  209-252.

\bibitem[Lando and Nielsen(2008)]{Lando.Nielsen2008} \rien Lando, D., Nielsen, M. S. (2008) \it Correlation in corporate defaults: contagion or conditional independence, \rm working paper, Copenhagen Business School.

\bibitem[Laurent, Cousin and Fermanian(2007)]{Laurent} \rien  Laurent, J.P., Cousin, A., Fermanian, J.D
 (2007) \it Hedging default risk for CDOs in Markovian contagion models, \rm{ Quantitative Finance}, Forthcoming.

\bibitem[Lopatin and Misirpashaev(2007)]{Lopatin.Misirpashaev2007} \rien  Lopatin, A.V., Misirpashaev, T. (2007) \it Two-dimensional Markovian model for dynamics of aggregate credit loss, \rm working paper, NumeriX.



\bibitem[R\"osch, Winterfeldt(2008)]{Rosch} \rien R\"osch, D., Winterfeldt, B. (2008) \it Estimating credit contagion in a standard factor model, \rm Risk, August 2008, S. 78-82.



\bibitem[Sakata, Hisakado and Mori(2007)]{SakataHisakadoMori}\rien
Sakata, A., Hisakado, M., Mori, S. (2007) \it Infectious Default
Model with Recovery and Continuous Limits, \rm Journal of the
Physical Society of Japan, Vol. 76, No. 5.


\bibitem[Sch{\"o}nbucher(2006)]{Schonbucher2006} \rien Sch{\"o}nbucher, P.J. (2006) \it Portfolio losses and the term-structure of loss transition rates: a new methodology for the pricing of portfolio credit derivatives,  \rm working paper, ETH Z{\"u}rich.

\bibitem[Sch\"onbucher and Schubert(2001)]{Schonbucher} \rien Sch\"onbucher, P.,
Schubert, D. (2001) \it Copula dependent default risk in intensity models,  \rm
Working paper, Bonn University.


\bibitem[Van der Voort(2006)]{Voort2006} \rien Van der Voort, M. (2006) \it An implied loss model, \rm
working paper, ABN Amro and Erasmus University.

\bibitem[Yu(2007)]{Yu} \rien Yu, F. (2007) \it Correlated defaults in intensity-based models, \rm
Mathematical Finance 17 (2), 155-173.
\end{thebibliography}
\end{document}